\documentclass[12pt]{article}
\pdfoutput=1

\RequirePackage{calc}
\RequirePackage{natbib}
\RequirePackage[american]{babel}

\usepackage{comment}
\usepackage{amsmath}
\usepackage{amsfonts}
\usepackage{graphicx}
\usepackage{float}
\usepackage{tikz}
\usetikzlibrary{shapes.geometric}
\usepackage{algorithm}
\usepackage{algorithmic}
\usepackage{url}
\usepackage{ulem}

\begin{document}

\begin{center}
  {\LARGE Phylogenetic Stochastic Mapping without Matrix Exponentiation}\\
  \vspace{0.2cm}
  {
  Jan Irvahn$^1$  and Vladimir N. Minin$^{1,2,*}$\\ 
   $^1$Department of Statistics and $^2$Department of Biology\\
      University of Washington, Seattle\\
    $^{*}$corresponding author: \url{vminin@uw.edu}
  }
\end{center}

\vspace{0cm}

\begin{abstract}

Phylogenetic stochastic mapping is a method for reconstructing the history of trait changes on a phylogenetic tree relating species/organisms carrying the trait. State-of-the-art methods assume that the trait evolves according to a continuous-time Markov chain (CTMC) and work well for small state spaces.  The computations slow down considerably for larger state spaces (e.g.\ space of codons), because current methodology relies on exponentiating CTMC infinitesimal rate matrices --- an operation whose computational complexity grows as the size of the CTMC state space cubed. In this work, we introduce a new approach, based on a CTMC technique called uniformization, that does not use matrix exponentiation for phylogenetic stochastic mapping. Our method is based on a new Markov chain Monte Carlo (MCMC) algorithm that targets the distribution of trait histories conditional on the trait data observed at the tips of the tree. 
The computational complexity of our MCMC method grows as the size of the CTMC state space squared. Moreover, in contrast to competing matrix exponentiation methods, if the rate matrix is sparse, we can leverage this sparsity and increase the computational efficiency of our algorithm further. Using simulated data, we illustrate advantages of our MCMC algorithm and investigate how large the state space needs to be for our method to outperform matrix exponentiation approaches. We show that even on the moderately large state space of codons our MCMC method can be significantly faster than currently used matrix exponentiation methods. 

\end{abstract}

\section{Introduction}

Phylogenetic stochastic mapping aims at reconstructing the history of trait changes on a phylogenetic tree that describes evolutionary relationships among organisms of interest. Such trait mapping on phylogenies has become a key element in computational evolutionary biology analyses. Stochastic mapping has been used successfully to enable computational analyses of complex models of protein evolution \citep{rodrigue2008uniformization,Rodrigue2010}, to reconstruct geographical movements of ancestral populations \citep{pereira2007mitochondrial,Lemey2009bayesian}, and to test hypotheses about morphological trait evolution \citep{Huelsenbeck2003vg,renner2007evolution}. Another testimony to the usefulness of stochastic mapping is the fact that this relatively new method has already been implemented in multiple widely used software packages: SIMMAP \citep{Bollback2006ju}, PhyloBayes \citep{Lartillot2009gb}, Bio++ libraries \citep{Gueguen2013}, and BEAST \citep{Drummond2012cs}.  Despite all these successes of stochastic mapping, this technique remains computationally challenging when the number of states that a trait can assume is large. Here, we present a new phylogenetic stochastic mapping algorithm that scales well with the size of the state space. 
\par
Stochastic mapping, initially developed by \citet{nielsen2002mapping} and subsequently refined by others \citep{lartillot2006conjugate,Hobolth2008}, assumes that discrete traits of interest evolve according to a continuous-time Markov chain (CTMC). Random sampling of evolutionary histories, conditional on the observed data, is accomplished by an algorithm akin to the forward filtering-backward sampling algorithm for hidden Markov models (HMMs) \citep{Scott2002}. However, since stochastic mapping operates in continuous-time, all current stochastic mapping algorithms require computing CTMC transition probabilities via matrix exponentiation --- a time consuming and potentially numerically unstable operation, when the CTMC state space grows large. \citet{de2010rapid} recognized the computational burden of the existing techniques and developed a faster, but approximate, stochastic mapping method based on time-discretization. We propose an alternative, {\it exact} stochastic mapping algorithm that relies on recent developments in the continuous-time HMM literature.
\par
\citet{rao2012fast} used a CTMC technique called uniformization to develop a method for sampling hidden trajectories in continuous time HMMs. The use of uniformization in this context is not new, but all previous methods produced independent samples of hidden trajectories with the help of matrix exponentiation --- an operation with algorithmic complexity $\mathcal{O}(s^3)$, where $s$ is the size of the CTMC state space \citep{Fearnhead2006}. \citet{rao2012fast} constructed a Markov chain Monte Carlo (MCMC) algorithm targeting the posterior distribution of hidden trajectories. Their new method eliminates the need for matrix exponentiation and results in an algorithm with complexity $\mathcal{O}(s^2)$. Moreover, the method of \citet{rao2012fast} can further increase its computational efficiency by taking advantage of sparsity of the CTMC rate matrix. Here, we take the method of \citet{rao2012fast} and extend it to phylogenetic stochastic mapping.  
\par
As in the original method of \citet{rao2012fast}, our new stochastic mapping method must pay a price for bypassing the matrix exponentiation step. The cost of the improved algorithmic complexity is the replacement of Monte Carlo in the state-of-the-art stochastic mapping with MCMC. Since Monte Carlo, if practical, is generally preferable to MCMC, it is not immediately clear that our new algorithm should be an improvement on the original method in all situations. We perform an extensive simulation study,  comparing performance of our new MCMC method with a matrix exponentiation method for different sizes of the state space. We conclude that, after accounting for dependence of trait history samples, our new MCMC algorithm can outperform existing approaches even on only moderately large state spaces ($s \sim 30$). We demonstrate additional computational efficiency of our algorithm when taking advantage of sparsity of the CTMC rate matrix. Since we suspect that our new method can speed up computations during studies of protein evolution, we examine in detail a standard GY94 codon substitution model ($s=61$) \citep{goldman1994codon}. We show that our new method can reduce computing times of state-of-the-art stochastic mapping by at least of factor of ten when working with this model. The last finding is important, because state-of-the-art statistical methods based on codon models often grind to a halt when applied to large datasets \citep{Valle2014}.


\section{Methods}

\subsection{CTMC model of evolution}


We start with a trait of interest, $X(t)$, and a rooted phylogenetic tree with $n$ tips and $2n-2$ branches. We assume that the phylogeny and its branch lengths, $\boldsymbol{\beta} = (\beta_1,\dots,\beta_{2n-2})$, are known to us. The trait can be in one of $s$ distinct states, $\{1,\dots,s\}$, at any particular place on the tree. We follow standard phylogenetics practice and assume that the trait evolves along the phylogenetic tree by the following stochastic process. First, a trait value is drawn at the root of the tree from an initial distribution $\boldsymbol{\pi} = (\pi_1,\dots,\pi_s)$. Next, starting at the root state, we use a CTMC with an infinitesimal $s\times s$ rate matrix $\mathbf{Q} = \{q_{kh}\}$ to produce two \textit{independent} CTMC trajectories along the branches leading to the two immediate descendants of the root node. After generating a CTMC trajectory along a branch we necessarily have generated a state for the child node of the branch. The procedure proceeds recursively by conditioning on a parent node and evolving the same CTMC independently along the two branches leading to the parent's children nodes. The random process stops when we reach the tips --- nodes that have no descendants. Trait states at the tips of the tree are observed, while the trait values everywhere else on the tree are considered missing data. We collect the observed data into a vector $\mathbf{y}$.


A substitution history for a phylogenetic tree is the complete list of transition events (CTMC jumps), including the time of each event (location on the tree) and the type of the transition event (e.g., $2\rightarrow 1$ transition). This state history can be encoded in a set of vectors, two vectors for each branch. Suppose branch $i$ has $n_i$ transitions so the full state history for branch $i$ can be described by a vector of state labels, ${\bf s}_i=(s_{i0},...,s_{in_i})$, 
and a vector of intertransition times, ${\bf t}_i=(t_{i0},...,t_{in_i})$.
Let $\mathcal{S}$ be the collection of all the ${\bf s}_i$ vectors and let $\mathcal{T}$ be the collection of all the ${\bf t}_i$ vectors, forming the full substitution history, $(\mathcal{S},\mathcal{T})$. See plot 3 in Figure~\ref{algorithm} for a substitution history example. The tree in Figure~\ref{algorithm} has four branches so the collection of state labels is $\mathcal{S}=\{{\bf s}_1,{\bf s}_2,{\bf s}_3,{\bf s}_4\}$, where ${\bf s}_1=(1)$, ${\bf s}_2=(1,2)$, ${\bf s}_3=(3)$, and  ${\bf s}_4=(3,1)$. The collection of intertransition times is $\mathcal{T}=\{{\bf t}_1,{\bf t}_2,{\bf t}_3,{\bf t}_4\}$, where ${\bf t}_1=(3.2)$, ${\bf t}_2=(0.64,2.56)$, ${\bf t}_3=(8)$, and  ${\bf t}_4=(2.4,2.4)$.

The goal of stochastic mapping is to be able to compute properties of the distribution of the substitution history of a phylogenetic tree conditional on the observed states at the tips of the tree, $\text{p}(\mathcal{S},\mathcal{T}|{\bf y})$.

\subsection{Nielsen's Monte Carlo Sampler}
\citet{nielsen2002mapping} proposed the basic framework that state-of-the-art phylogenetic stochastic mapping currently uses. His approach samples directly from the conditional distribution, $\text{p}(\mathcal{S},\mathcal{T}|{\bf y})$, in three steps. First, one calculates partial likelihoods using Felsenstein's algorithm \citep{felsenstein1981evolutionary}. The partial likelihood matrix records the likelihood of the observed data at the tips of the tree beneath each node after conditioning on the state of said node. This requires calculating transition probabilities for each branch via matrix exponentiation. Second, one recursively samples internal node states, starting from the root of the tree. Third, one draws realizations of CTMC trajectories on each branch conditional on the sampled states at the branch's parent and child nodes. The last step can be accomplished by multiple algorithms reviewed in \citep{hobolth2009simulation}. In order to avoid matrix exponentiation while approximating $\text{p}(\mathcal{S},\mathcal{T}|{\bf y})$ we will need a data augmentation technique called uniformization.

\subsection{CTMC Uniformization}

An alternative way to describe the CTMC model of evolution on a phylogenetic tree uses a homogenous Poisson process coupled with a discrete time Markov chain (DTMC) that is independent from the Poisson process. The intensity of the homogenous Poisson process, $\Omega$, must be greater than the largest rate of leaving a state, $\text{max}_k|q_{kk}|$. The generative process that produces a substitution history on a phylogenetic tree first samples the total number of DTMC transitions over the tree, $N$, drawn from a Poisson distribution with mean equal to $\Omega \sum_{i=1}^{2n-2}\beta_i$ --- the product of the Poisson intensity and the sum of all the branch lengths. The locations/times of the $N$ transitions are then distributed uniformly at random across all the branches of the tree. These transition time points separate each branch into segments. The intertransition times (the length of each segment) for branch $i$ compose the vector, ${\bf w}_i$, where the sum of elements of this vector equal the branch length $\beta_i$. The state of each segment evolves according to a DTMC with transition probability matrix $\mathbf{B} = \{b_{kh}\}$ satisfying ${\bf B}={\bf I}+{\bf Q}/\Omega$. 

Again, the uniformized generative process samples a state at the root of the tree and works down the tree sampling the state of each branch segment sequentially. Conditional on the previous/ancestral segment being in state $k$, we sample the current segment's state from a multinomial distribution with probabilities $(b_{k1}, \dots, b_{ks})$. The states of each segment of branch $i$ compose the vector, ${\bf v}_i$. It is important to note that the stochastic transition matrix ${\bf B}$ allows the DTMC to transition from state $k$ to state $k$, i.e., self transitions are allowed. Intuitively, the dominating homogenous Poisson process produces more transition events (on average) than we would expect under the CTMC model of evolution. The DTMC allows some of the transitions generated by the Poisson process to be self transitions so that the remaining ``real'' transitions and times between them yield the exact CTMC trajectories we desire \citep{jensen1953markoff}.

An augmented substitution history of a phylogenetic tree encodes all the information in $(\mathcal{S},\mathcal{T})$ and adds virtual jump times as seen in plot 4 of Figure \ref{algorithm}. The notation describing an augmented substitution history is similar to the notation used to describe a substitution history. One branch is fully described by two vectors. Let branch $i$ have $m_i$ jumps (real and virtual) and again, ${\bf v}_i =(v_{i0},...,v_{im_i})$ is a vector of state labels, ${\bf w}_i =(w_{i0},...,w_{im_i})$ is a vector of intertransition times,
$\mathcal{V}$ is the collection of all the ${\bf v}_i$ vectors, and $\mathcal{W}$ is the collection of all the ${\bf w}_i$ vectors. The augmented state history is $(\mathcal{V},\mathcal{W})$. The tree in plot 4 of Figure~\ref{algorithm} has four branches so the collection of state labels is $\mathcal{V}=\{{\bf v}_1,{\bf v}_2,{\bf v}_3,{\bf v}_4\}$, where ${\bf v}_1=(1,1)$, ${\bf v}_2=(1,2)$, ${\bf v}_3=(3,3)$, and  ${\bf v}_4=(3,1)$. The collection of intertransition times is $\mathcal{W}=\{{\bf w}_1,{\bf w}_2,{\bf w}_3,{\bf w}_4\}$, where ${\bf w}_1=(1.6,1.6)$, ${\bf w}_2=(0.64,2.56)$, ${\bf w}_3=(7,1)$, and ${\bf w}_4=(2.4,2.4)$. The locations/times of each virtual jump on branch $i$ is represented by a vector ${\bf u}_i=(u_{i1},..,u_{i(m_i-n_i)})$. For example, the distance from the parent node of branch $i$ to the $d^{\text{th}}$ virtual jump is $u_{id}$. The collection of the ${\bf u}_i$ vectors, fully determined by $(\mathcal{V},\mathcal{W})$, is denoted by $\mathcal{U}$. In plot 4 of Figure~\ref{algorithm} the collection of virtual jump times is $\mathcal{U}=\{{\bf u}_1,{\bf u}_2,{\bf u}_3,{\bf u}_4\}$ where ${\bf u}_1=(1.6)$, ${\bf u}_2=()$, ${\bf u}_3=(7)$, and ${\bf u}_4=()$.

\subsection{New MCMC Sampler}
Equipped with notation describing the CTMC model of evolution and a companion uniformization process, we now turn our attention to making inference about a phylogenetic tree state history conditional on observed data. We investigate the situation where the tree topology is fixed, branch lengths are fixed, and the rate matrix parameters are all known and fixed. The goal is to construct an ergodic Markov chain on the state space of augmented substitution histories with the stationary distribution $\text{p}(\mathcal{V},\mathcal{W}|{\bf y})$. 
\par
Our MCMC sampler uses two Markov kernels to create a Markov chain whose stationary distribution is $\text{p}(\mathcal{V},\mathcal{W}|{\bf y})$. The first kernel samples from $\text{p}(\mathcal{V}|\mathcal{W},{\bf y})$ --- the distribution of states on the tree conditional on tip states and the jump locations on each branch.
A Markov chain that sequentially draws from this full conditional has $\text{p}(\mathcal{V},\mathcal{W}|{\bf y})$ as its stationary distribution. This kernel alone is not ergodic because the set of transition times, $\mathcal{W}$, is not updated. To create an ergodic Markov chain we introduce a second Markov kernel to sample from 
$\text{p}(\mathcal{U}|\mathcal{S},\mathcal{T},{\bf y})$ --- the distribution of virtual transitions conditional on the substitution history.
Again, drawing from the full conditional of $\mathcal{U}$ ensures that $\text{p}(\mathcal{V},\mathcal{W}|{\bf y})$ is a stationary distribution of this kernel. This kernel alone is not ergodic either but when the two kernels are combined we create an ergodic Markov chain with the desired stationary distribution. In general, it takes two sequential applications of the above kernels before the probability density of transitioning between two arbitrary augmented substitution histories becomes nonzero. 

\subsubsection{Sampling States from $\text{p}(\mathcal{V}|\mathcal{W},{\bf y})$}
Our strategy for sampling states $\mathcal{V}$ is to make a draw from the full conditional of internal node states and then to sample the states along each branch conditional on 
the branch's parent and child nodes. It is useful to remember that when conditioning on the number of virtual and real jumps, and locations of these jumps on the tree, 
our data generating process becomes a DTMC with transition probability matrix $\mathbf{B}$ and a known number of transitions on each branch. Alternatively, we can think of
the trait evolving along each branch $i$ according a branch-specific DTMC with transition probability matrix $\mathbf{B}^{m_i}$, where $m_i$ is the number of transitions 
on branch $i$. This is similar to representing a regular, non-uniformized, CTMC model as a collection of branch-specific DTMCs with transition probability matrices 
$\mathbf{P}(\beta_1),\dots, \mathbf{P}(\beta_{2n-2})$. This analogy allows us to use standard algorithms for sampling internal node states on a phylogenetic tree by
replacing in these algorithms $\mathbf{P}(\beta_i)$ with $\mathbf{B}^{m_i}$ for all $i = 1,\dots,2n-2$. For completeness, we make this substitution explicit below.
\par
We start by using Felsenstein's algorithm to compute a partial likelihood matrix ${\bf L} = \{l_{jk}\}$, where $l_{jk}$ is the probability of the observed tip states below node $j$ given that node $j$ is in state $k$
\citep{felsenstein1981evolutionary}. 
The matrix $\textbf{L}$ has $(2n-1)$ rows and $s$ columns because there are $(2n-1)$ nodes (including the tips) and there are $s$ states. Starting at the tips, we work our way up the tree calculating partial likelihoods at internal nodes as we go. 
We need to calculate the partial likelihood at both child nodes before calculating the partial likelihood at a parent node because of the recursive nature of the Felsenstein algorithm. 
The algorithm is initialized by setting each row corresponding to a tip node to zeros everywhere except for the column corresponding to the observed state of that tip. The matrix value at this entry is set to 1. 
Next, we calculate the partial likelihoods for all the internal nodes. Suppose branch $i$ connecting parent node $p$ to child node $c$ has $m_i$ jumps so the probability transition matrix for branch $i$ is 
${\bf E}(m_i)\equiv {\bf B}^{m_i}$. The probability of transitioning from state $h$ to state $k$ along branch $i$ is $\text{e}(i)_{hk}$, the $(h,k)^{\text{th}}$ element of ${\bf E}(m_i)$. We refer to the state of node $j$ as $y_j$. The probability of observing the tip states below node $c$ conditional on node $p$ being in state $h$ is
\begin{align*}
g_{pch}&=\sum_{k=1}^s \text{Pr}(y_c=k|y_p=h)l_{ck} =\left({\bf e}(m_i)_{h-}\right)(\textbf{l}_{c-})^T,
\end{align*}
where $\textbf{l}_{c-} = (l_{c1},\dots, l_{cs})$. If node $c$ is a tip then conditioning on the tip states below $c$ is the same as conditioning on the state of tip $c$. We combine the probabilities, $g_{pch}$, for each state $h$ into a single vector, ${\bf g}_{pc-}$, and then create the same type of vector for the second branch below node $p$, ${\bf g}_{pd-}$. Element wise multiplication of the two vectors yields the vector of partial likelihoods for node $p$:
\begin{align*}
(\textbf{l}_{p-})^T&={\bf g}_{pc-}*{\bf g}_{pd-}.
\end{align*}
After working our way up the tree we have the matrix of partial likelihoods, $\textbf{L}$. 

\paragraph{Sampling internal node states}
Starting at the root we work our way down the tree sampling the states of internal nodes conditional on tip states, the number of jumps on each branch, and previously sampled internal node states. The prior probability that the root is in state $k$ is the $k^{\text{th}}$ element of the probability vector $\boldsymbol{\pi}$. The probability that the root is in state $k$ given the states of all the tip nodes is,
\begin{align*}
\text{Pr}(y_{\text{root}}=k|\bf{y}) &= \frac{\text{Pr}(y_{\text{root}}=k \hspace{1mm}\& \ {\bf y})}{\text{Pr}({\bf y})}
=\frac{\text{Pr}(y_{\text{root}}=k)\text{Pr}({\bf y}|y_{\text{root}}=k)}{\sum_{h=1}^s \text{Pr}(y_{\text{root}}=h)\text{Pr}({\bf y}|y_{\text{root}}=h)}
=\frac{\pi_k l_{(\text{root }k)}}{\textbf{l}_{(\text{root -})} \boldsymbol{\pi}}.
\end{align*}
Once we calculate the probability of the root being in each possible state we sample the state of the root from the multinomial distribution with probabilities we just computed.
Next, we sample all non-root, non-tip nodes. 
Without loss of generality, let us consider node $c$ connected to its parent node, node $p$, by branch $i$. 
Suppose node $p$'s previously sampled state is $h$ and the number of jumps on branch $i$ is $m_i$. 
The vector containing observed tip states at the eventual descendants of node $c$ is ${\bf d}_c$. 
The probability that node $c$ is in state $k$ given node $p$ is in state $h$ and given the state of the tips below node $c$ is
\begin{align}
\text{Pr}(y_c = k|y_p=h \hspace{1mm}\&\  {\bf d}_c) =\frac{\text{Pr}(y_c=k|y_p=h)\text{Pr}({\bf d}_c|y_c=k)}{\sum_{k=1}^s \text{Pr}(y_c=k|y_p=h)\text{Pr}({\bf d}_c|y_c=k)} =\frac{\text{e}(m_i)_{hk}\text{l}_{ck}}{{\bf e}(m_i)_{h-}(\textbf{l}_{c-})^T}.
\label{int_node_post_prob}
\end{align}
Starting with the root we can now work our way down the tree sampling the states of each node from the multinomial distributions with probabilities we just described. 
Equation (\ref{int_node_post_prob}) may suggest that sampling internal nodes has algorithmic complexity 
$\mathcal{O}(s^3)$, because raising a matrix to a power requires $\mathcal{O}(s^3)$ multiplications. It is important to note that we never need to calculate ${\bf B}^m$ as we only need ${\bf B}^m \times \text{vector}$, which requires $\mathcal{O}(s^2)$ multiplications. 

\paragraph{Sampling branch states}

We sample the states on each branch separately, conditioning both on previously sampled internal nodes states and on the number of transitions on each branch. Conditioning on the internal node states means the starting and ending state of each branch are set so we only sample internal segments of the branches. Conditioning on the number of transitions on a branch means we are sampling states of the discrete time Markov chain with transition matrix ${\bf B}$. Suppose branch $i$ starts in state $v_{i0}$ and ends in state $v_{im_i}$ (or $y_c$). We sample each segment of the branch in turn, starting with the second segment because the first segment has to be in the same state as the parent node of the branch. The state of each segment is sampled conditional on the state of the previous segment, the number of transitions until the end of the branch, and the ending state of the branch, $v_{im_i}=y_c$. The state of the $d^{\text{th}}$ segement is sampled from a multinomial distribution with probabilities calculated according to the following formula: 
\begin{align*}
& \text{Pr}(v_{id}=k|v_{i(d-1)}=h,v_{im_i}=y_c)=\frac{\text{Pr}(v_{i(d-1)}=h,v_{id}=k,v_{im_i}=y_c)}{\text{Pr}(v_{i(d-1)}=h,v_{im_i}=y_c)}\\
&=\frac{\text{Pr}(v_{i(d-1)}=h)\text{Pr}(v_{id}=k|v_{i(d-1)}=h)\text{Pr}(v_{im_i}=y_c|v_{id}=k)}{\text{Pr}(v_{i(d-1)}=h)\text{Pr}(v_{im_i}=y_c|v_{i(d-1)}=h)}
=\frac{b_{hk}\text{e}(m_i-d)_{ky_c}}{\text{e}(m_i-d+1)_{hy_c}}.
\end{align*}
After sampling the states along each branch we have completed one cycle through the first Markov kernel by sampling from 
$\text{p}(\mathcal{V}|\mathcal{W},{\bf y})$. The second Markov kernel requires us to sample virtual transitions 
conditional on the current substitution history (not augmented by virtual jumps).

\subsubsection{Sampling Virtual Jumps from $\text{p}(\mathcal{U}|\mathcal{S},\mathcal{T},{\bf y})$}

After sampling the states on each branch, $\mathcal{V}$, we resample virtual jumps, $\mathcal{U}$, on each branch separately. Without loss of generality  consider a branch with a newly sampled substitution history, $({\bf s},{\bf t})$, which is the augmented substitution history with all the virtual jumps removed. Suppose the branch contains $n$ real transitions. Resampling virtual jumps for the branch involves resampling virtual jumps for each of the $n+1$ segments of the branch separately. To sample the a$^{\text{th}}$ segment of the branch we need to sample the number of virtual jumps, $\mu_{a}$, and the locations of these virtual jumps. After sampling virtual jumps for each of the $n+1$ branch segments we have $m$ transitions total, both real and virtual, so that $m=n+\sum_{a=0}^{n}\mu_a$.  Careful examination of the likelihood of the dominating homogenous Poisson process for a single branch of the tree allows us to derive the distribution of virtual jumps conditional on the substitution history of the branch. 

Suppose there are $m$ jumps along a branch including real transitions and virtual transitions. Let $v_d$ be the state of the chain after the $d^{\text{th}}$ transition and let $\pi'_{v_0}$ be the probability that the branch starts in state $v_0$. The density of the augmented substitution history is
\begin{equation}
\text{p}({\bf v},{\bf w})=\pi'_{v_0}\frac{e^{-\Omega t}(\Omega t)^m}{m!}\frac{m!}{t^m}\prod_{d=1}^m B_{v_{d-1},v_{d}}.
\label{aug_sub_hist_like}
\end{equation}
The density as written above has four parts, the probability of starting in state $v_{0}=s_0$, the probability of $m$ transition points, the density of the locations of $m$ unordered points conditional on there being $m$ points, and the probability of each transition in a discrete time Markov chain with transition matrix ${\bf B}$. 

The density of the augmented substitution history of one branch, $\text{p}({\bf v},{\bf w})$, can be rewritten as $\text{p}({\bf u},{\bf s},{\bf t})$, because the substitution history, $({\bf s},{\bf t})$ combined with the virtual jump locations, ${\bf u}$, form the augmented substitution history. To derive the full conditional for $\mathbf{u}$, we follow \citet{rao2012fast} and rewrite density (\ref{aug_sub_hist_like}) as follows:

\begin{align*}
\text{p}({\bf u},{\bf s},{\bf t}) =
\text{p}({\bf v},{\bf w})
=\prod_{a=0}^n \left(r_a^{\mu_a}e^{-r_at_a}\right)\pi'_{s_0}\left(\prod_{z=1}^{n}|q_{s_{z-1}}|e^{q_{s_{z-1}}t_{z-1}}\frac{q_{s_{z-1}s_{z}}}{|q_{s_{z-1}}|}\right)e^{q_{s_{n}}t_{n}}
=\prod_{a=0}^n \left(r_a^{\mu_a}e^{-r_at_a}\right)\text{p}({\bf s},{\bf t}),
\end{align*}
where $q_{s_a}\equiv q_{s_as_a}$ and $r_a=\Omega+q_{s_a}$. Therefore,
\begin{equation}
\text{p}({\bf u}|{\bf s},{\bf t}, \mathbf{y}) = \text{p}({\bf u}|{\bf s},{\bf t})= \frac{\text{p}({\bf u},{\bf s},{\bf t})}{\text{p}({\bf s},{\bf t})} = \prod_{a=0}^n \left(r_a^{\mu_a}e^{-r_at_a}\right) = 
\prod_{a=0}^n \frac{e^{-r_at_a}(r_at_a)^{\mu_a}}{\mu_a!}\frac{\mu_a!}{t_a^{\mu_a}}.
\label{self_jumps_full_cond}
\end{equation}

\begin{figure}
\begin{center}
\includegraphics[width=\textwidth]{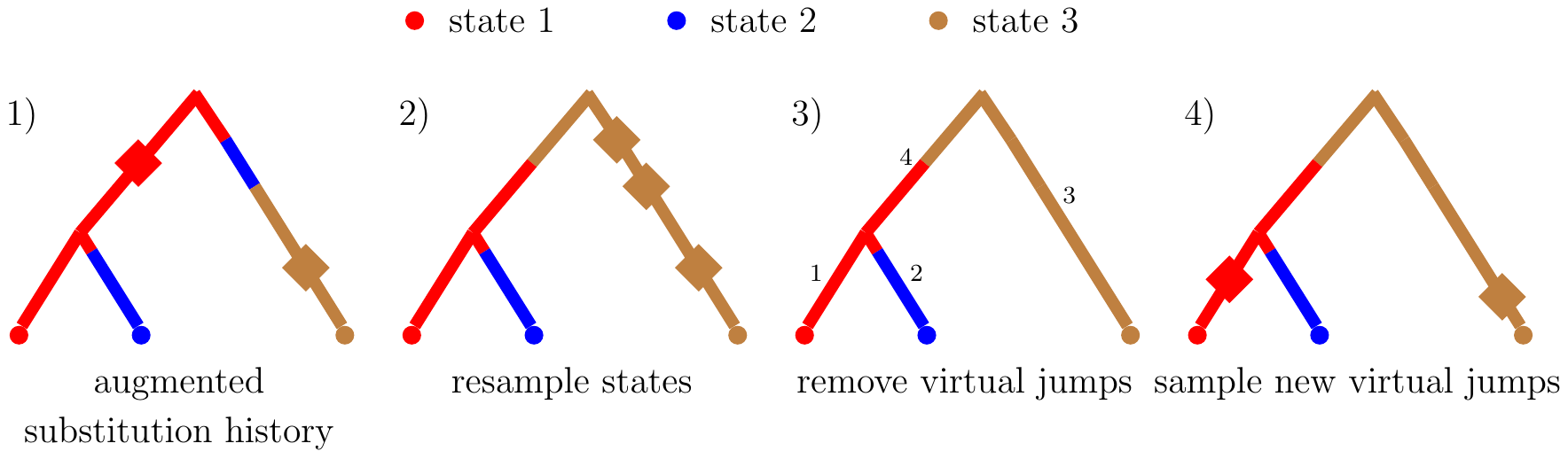}
\end{center}
\caption{An example of applying the two Markov kernels of our MCMC sampler to an augmented substitution history. The diamonds represent virtual transitions.  1) shows an initial augmented substitution history; 2) shows the substitution history after resampling states on the phylogeny conditional on tip node states and the transition points (both real and virtual); 3) shows the substitution history seen in 2) with no virtual jumps; 4) shows the augmented substitution history after resampling virtual jumps conditional on the substitution history seen in 3). The transition from 1) to 2) shows the effect of the first Markov kernel, sampling from $\text{p}(\mathcal{V}|\mathcal{W},{\bf y})$. The transition from 3) to 4) shows the effect of the second Markov kernel, sampling from $\text{p}(\mathcal{U}|\mathcal{S},\mathcal{T})$.}
\label{algorithm}
\end{figure}

The full conditional density (\ref{self_jumps_full_cond}) is a density of an inhomogenous Poisson process with intensity $r(t)=\Omega+q_{X(t)}$. This intensity is piecewise constant so we can add self transition locations/times to a branch segment in state $s_a$ by drawing a realization of a homogenous Poisson process with rate $r_a = \Omega + q_{s_a}$. More specifically, we sample the number of self transitions, $\mu_a$, on this segment by sampling from a Poisson distribution with mean $r_at_a$ and then distributing the locations/times of the $\mu_a$ self transitions uniformly at random across the segment. This procedure is repeated independently for all segments on all branches of the phylogenetic tree, concluding our MCMC development, summarized in Algorithm~\ref{mcmc_algorithm} and illustrated in Figure~\ref{algorithm}.

\begin{algorithm}[H]
 \caption{MCMC for phylogenetic stochastic mapping}
\begin{algorithmic} [1]
\STATE Start with an augmented substitution history, $(\mathcal{V}_0,\mathcal{W}_0)$ 
 \FOR{$\gamma \in \{1,3,5,\dots,2N-1\}$}
  \STATE sample from $\text{p}(\mathcal{V}_{\gamma}|\mathcal{W}_{\gamma-1},{\bf y})$ \hspace{2mm}producing a new substitution history $(\mathcal{S}_{\gamma},\mathcal{T}_{\gamma})$
  \renewcommand{\labelenumi}{(\roman{enumi})}
  \begin{enumerate}
 \item sample internal node states conditional on $\mathbf{y}$ and the  number of jumps on each branch
 \begin{enumerate}
 \item starting at the tips work up the tree calculating partial likelihoods
 \item starting at the root work down the tree sampling internal node states
 \end{enumerate}
 \item sample segmental states conditional on end states and number of jumps
\end{enumerate}
\STATE sample from $\text{p}(\mathcal{U}_{\gamma+1}|\mathcal{S}_{\gamma},\mathcal{T}_{\gamma})$, producing
$(\mathcal{V}_{\gamma+1},\mathcal{W}_{\gamma+1})$
 \renewcommand{\labelenumi}{(\roman{enumi})}
  \begin{enumerate}
  \item remove virtual jumps
  \item sample virtual jumps conditional on substitution history
  \end{enumerate}
\ENDFOR
\RETURN $(\mathcal{V}_0,\mathcal{W}_0), (\mathcal{V}_2,\mathcal{W}_2),\dots, (\mathcal{V}_{2N},\mathcal{W}_{2N})$
\end{algorithmic}
\label{mcmc_algorithm}
\end{algorithm}

\section{Assessing Computational Efficiency}

\subsection{Algorithm Complexity}

State-of-the-art stochastic mapping approaches rely on exponentiating CTMC rate matrices, requiring $\mathcal{O}(s^3)$ operations. Our MCMC algorithm uses only matrix-by-vector multiplications, allowing us to accomplish the same task in $\mathcal{O}(s^2)$ operations. Moreover, if the CTMC rate matrix is sparse, the algorithmic complexity of our method
can go down further. For example, if $\mathbf{Q}$ is a tri-diagonal matrix, as in the birth-death CTMCs used to model evolution of gene family sizes \citep{Spencer2006}, 
then our MCMC achieves an algorithmic complexity of $\mathcal{O}(s)$. In contrast, even after disregarding the cost of matrix exponentiation, approaches relying on this operation 
require at least $\mathcal{O}(s^2)$ operations, because $e^{\mathbf{Q}t}$ is a dense matrix regardless of the sparsity of $\mathbf{Q}$.
However, since the number of matrix-by-vector multiplications is a random variable in our algorithm, the algorithmic complexity with respect to the state space size does not tell the whole story, prompting us to perform an empirical comparison of the two approaches in a set of simulation studies. 
In these simulation studies, we need to compare state-of-the-art Monte Carlo algorithms and our MCMC in a principled way, which we describe in the next subsection.

\subsection{Effective Sample Size}
When comparing timing results of our MCMC approach and a matrix exponentiation approach, we need to account for the fact that our MCMC algorithm produces correlated substitution histories. 
One standard way to compare computational efficiency of MCMC algorithms is by 
reporting CPU time divided by effective sample size (ESS), where ESS is a measure
of the autocorrelation in a stationary time series \citep{Holmes2006, Girolami2011}.
More formally, the ESS of a stationary time series of size $N$ with stationary distribution 
$\nu$ is an integer $N_{\text{eff}}$ such that $N_{\text{eff}}$ independent realizations from $\nu$ have the same sample variance as the sample variance of the time series. 
The ESS of a stationary time series of size $N$ is generally less than $N$ and is equal to 
$N$ if the time series consists of independent draws from $\nu$. 

\par
In MCMC literature, ESSs are usually calculated for model parameters, latent variables, and  the log-likelihood. Since we are fixing model parameters in this paper, 
we monitor ESSs for our latent variables --- augmented substitution history summaries --- and  $\log \text{p}(\mathcal{S},\mathcal{T})$ --- the log-density of the substitution history. 
Although the amount of time spent in each state over the entire tree and the numbers of transitions between each possible pair of states are sufficient statistics of
a fully observed CTMC \citep{Guttorp1995}, it is impractical to use all of these summaries for ESS calculations. This stems from the fact that we are interested in the parameter regimes 
under which we expect a small number of CTMC transitions over the entire tree. In such regimes, some of the states are never visited so the amount of time spent in these states is zero, 
which creates an impression that the MCMC is mixing poorly. To avoid this problem, 
we restrict our attention to the amount of time spent (over the entire tree) in each of the states that are \textit{observed} at the tips. Similarly,
we restrict our attention to transition counts between \textit{observed} tip states. Each of the univariate statistics, including the log-density of the substitution history, yields 
a potentially different ESS, which we calculate with the help of the R package \verb8coda8 \citep{coda2006}. We follow \citet{Girolami2011} and conservatively use the 
minimum of these univariate ESSs to normalize the CPU time of running our MCMC sampler. More specifically, in all our numerical experiments, 
we generate 10,000 substitution histories via both MCMC and matrix exponentiation methods and then multiply the CPU time of our MCMC sampler by 
$10,000/\min(\text{univariate ESSs})$.


\subsection{Matrix Exponentiation}

In all our simulations we compare timing results of our new MCMC approach with another CTMC uniformization approach that relies on matrix exponentiation \citep{lartillot2006conjugate}. 
For the matrix exponentiation approach we recalculate the partial likelihood matrix at each iteration, which involves re-exponentiating the rate matrix. 
We do so in order to learn how our MCMC method will compare to the matrix exponentiation method in situations where the parameters of the rate matrix are updated during a MCMC that targets the joint posterior of substitution histories and CTMC parameters \citep{lartillot2006conjugate, rodrigue2008uniformization}. 
Since matrix exponentiation is 
a potentially unstable operation \citep{moler1978nineteen}, we do not repeat it at each iteration in our simulations. Instead, we pre-compute an eigen 
decomposition of the CTMC rate matrix once, cache this decomposition and then use it to exponentiate $\mathbf{Q}$ at each iteration. Even though exponentiating
$\mathbf{Q}$ using its pre-computed eigen decomposition is an $\mathcal{O}(s^3)$ operation, our simulations do not fully mimic a more realistic procedure that repeatedly 
re-exponentiates the rate matrix. Skipping the eigen decomposition operation at each iteration of stochastic mapping increases computational efficiency of 
the matrix exponentiation method, making our timing comparisons conservative. 
\par
In one of our simulation studies, when we consider the effect of sparsity in the rate matrix, we depart from this matrix exponentiation regime. Instead of exponentiating the rate matrix at each iteration we exponentiate the rate matrix $\mathbf{Q}$ and compute the partial likelihood matrix once, 
sampling substitution histories at each iteration without recalculating branch-specific transition probabilities or partial likelihoods.  
We refer to this method as ``EXP once.'' 
We do not believe that our MCMC method is the most appropriate in this regime, but we are interested in how our new method compares to state-of-the-art 
methods when the calculations requiring $\mathcal{O}(s^3)$ operations were not involved.

\subsection{Implementation}
We have implemented our new MCMC approach in an R package \verb8phylomap8, available at \url{https://github.com/vnminin/phylomap}. The package also contains our implementation
of the matrix exponentiation-based uniformization method of \citet{lartillot2006conjugate}. We reused as much code as possible between these two stochastic mapping methods in order to minimize the impact
of implementation on our time comparison results. We coded all computationally intensive parts in C++ with the help of the \verb8Rcpp8 package \citep{Eddelbuettel2011}. We used 
the \verb8RcppArmadillo8 package to perform sparse matrix calculations \citep{Eddelbuettel2014}.

\section{Numerical Experiments}

\subsection{General Set Up}
We started all of our simulations by creating a random tree with 50 or 100 tips using the \verb8diversitree8 R package \citep{fitzjohn2012diversitree}. 
For each simulation that required the construction of a rate matrix, we set the transition rates between all state pairs to be identical. 
We then scaled the rate matrix for each tree so that the number of expected CTMC transitions per tree was either 2 or 6. 
These two values were intended to mimic slow and fast rates of evolution. 
Six expected transitions in molecular evolution settings is usually considered unreasonably high but six transitions (or more) is reasonable in other settings like phylogeography. 
For example, investigations of \citet{Lemey2009bayesian} into the geographical spread of human influenza H5N1 found on the order of 40 CTMC transitions on their phylogenies. 
To obtain each set of trait data we simulated one full state history after creating a tree and a rate matrix. 
We used this full state history as the starting augmented substitution history for our MCMC algorithm.
\par
To ensure our implementation of the matrix exponentiation approach properly sampled from $\text{p}(\mathcal{S},\mathcal{T}| \mathbf{y})$ and to ensure the stationary distribution of our MCMC approach was $\text{p}(\mathcal{V},\mathcal{W}| \mathbf{y})$, we compared distributions of univariate statistics produced by 
our implementations to the same distributions obtained by using \verb8diversitree8's implementation of phylogenetic stochastic mapping. 
We found that all implementations, including \verb8diversitree8's, appeared to produce the same distributions. Boxplots and histograms showing the results of our investigations can be found in Appendix A of the Supplementary Materials. 

\subsection{MCMC Convergence}
Although we have outlined a strategy for taking MCMC mixing into account via the normalization by ESS, we have not addressed possible problems with 
convergence of our MCMC. Examination of MCMC chains with an initial distribution different from the stationary distribution  showed very rapid convergence to
stationarity, as illustrated in Figure C-1 in the Supplementary Materials. Such rapid convergence is not surprising in light of the fact that we jointly 
update a large number of components in our MCMC state space without resorting to Metropolis-Hastings updates.

 \subsection{Effect of State Space Size}

\begin{figure}[!htbp]
\begin{center}
\includegraphics[width=.9\textwidth]{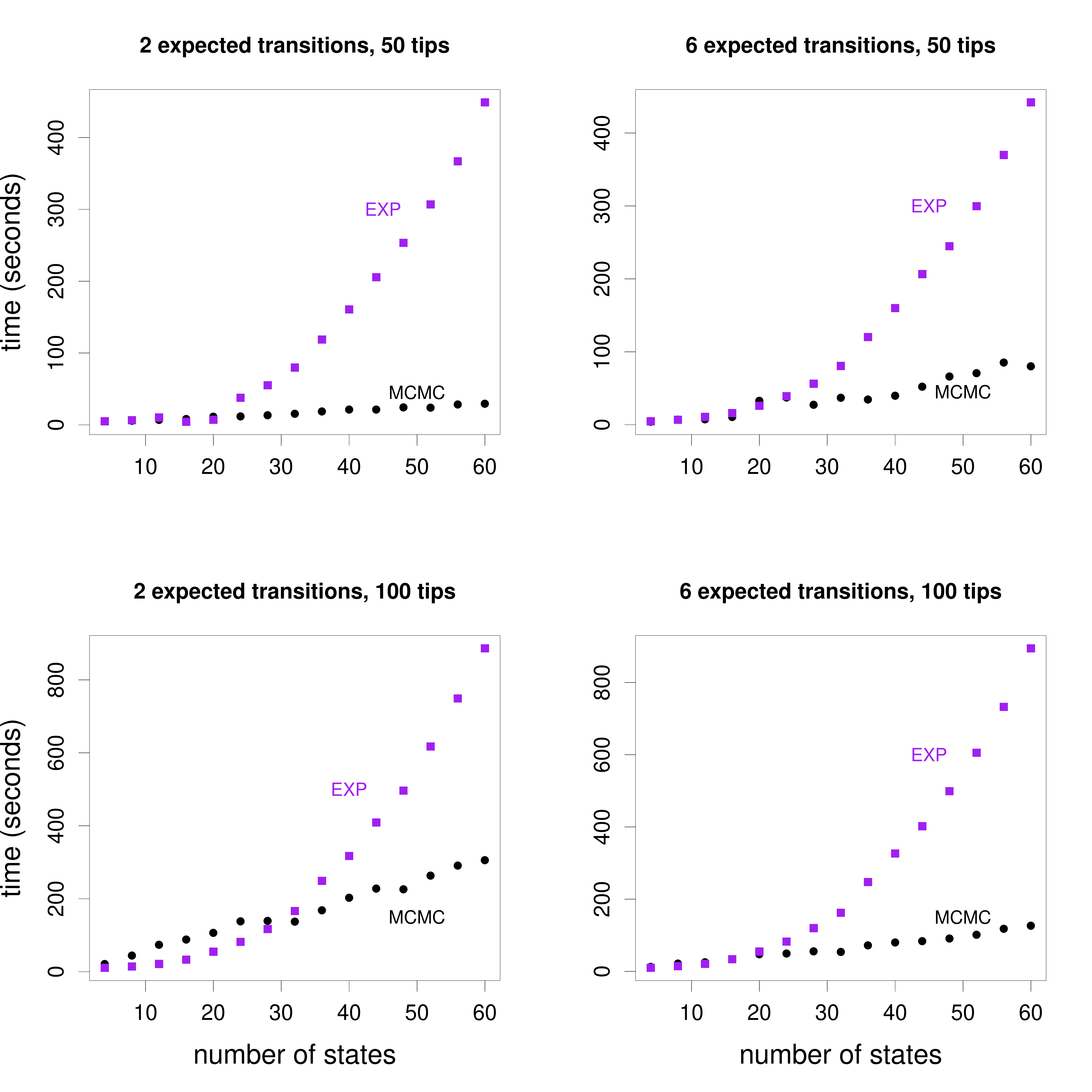}
\end{center}
\vspace{-0.5cm}
\caption{State space effect. 
All four plots show the amount of time required to obtain 10,000 effective samples as a function of the size of the state space for two methods, matrix exponentiation in purple squares 
and our MCMC sampler in black circles. The two plots in the top row show results for a randomly generated tree with 50 tips. The two plots in the bottom row show results for a randomly generated tree with 100 tips. The two plots in the left column show results for a rate matrix that was scaled to produce 2 expected transitions while the two plots in the right column show results for a rate matrix that was scaled to produce 6 expected transitions.}
\label{timeVstate}
\end{figure}

Our MCMC method scales more efficiently with the size of the state space than matrix exponentiation methods so we were first interested in comparing running times of the two approaches as the size of the CTMC state space increased.  In Figure \ref{timeVstate}, we show the amount of time it took the matrix exponentiation method to obtain 10,000 samples for different state space sizes and we show the amount of time it took our MCMC method to obtain an ESS of 10,000 for different state space sizes. The size of the state space varied between 4 states and 60 states. The tuning parameter, $\Omega$, was set to 0.2, ranging between 15 and 103 times larger than the largest rate of leaving a state. 

Figure \ref{timeVstate} contains timing results for four different scenarios. We considered two different rates of evolution corresponding to 2 expected transitions per tree and 6 expected transitions per tree and we considered two different tree tip counts, 50 and 100. The MCMC approach started to run faster than the matrix exponentiation approach when the size of the state space entered the 25 to 35 state range. At 60 states the MCMC approach was clearly faster in all four scenarios. For the senario involving 100 tips, 2 expected transtions, and 60 states the MCMC method was almost 3 times faster than the matrix exponentiation approach. For the scenario involving 50 tips, 2 expected transtions, and 60 states the MCMC method was about 15 times faster than the matrix exponentiation approach.
\begin{figure}[!htbp]
\begin{center}
\includegraphics[width=.9\textwidth]{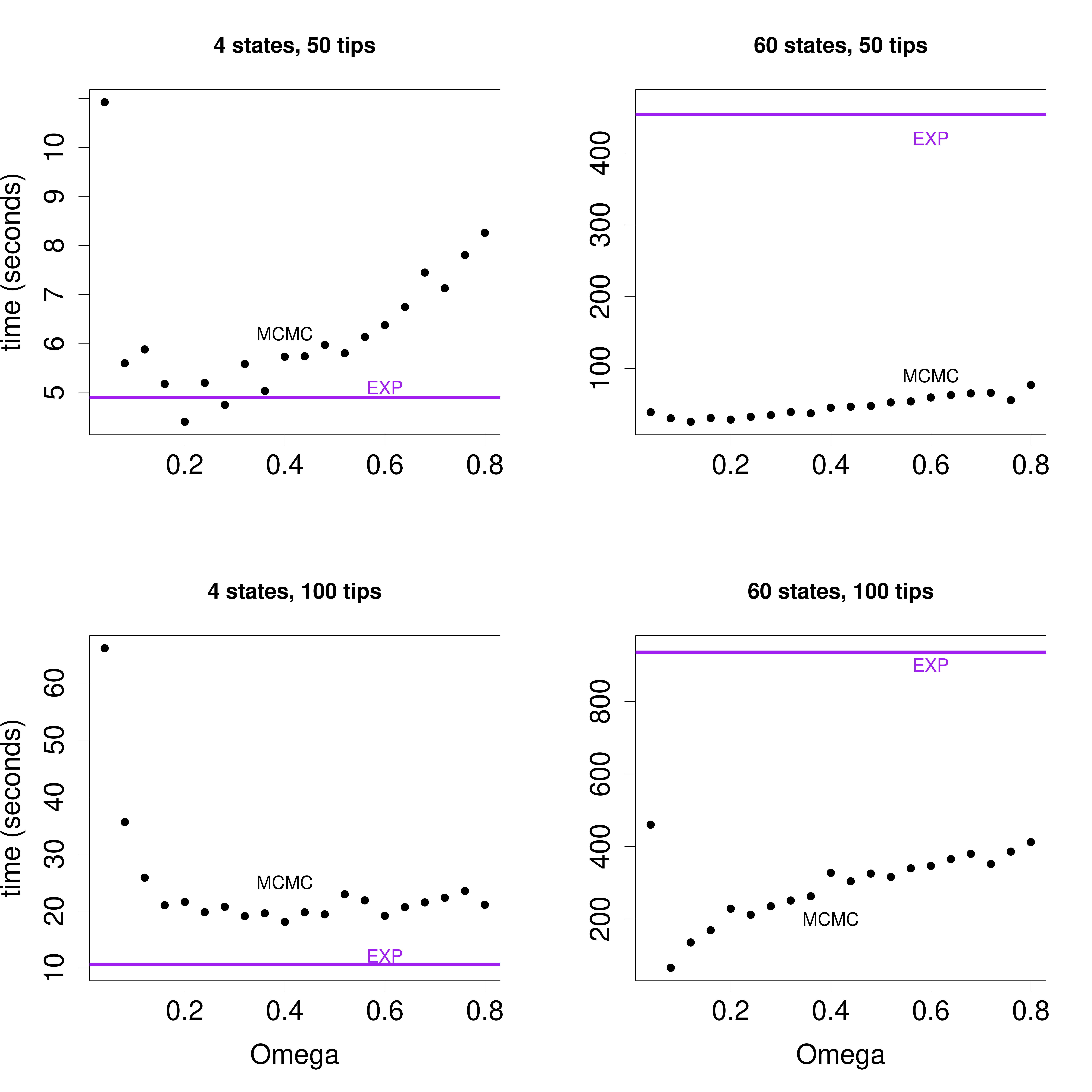}
\end{center}
\caption{Time to obtain 10,000 effective samples as a function of the dominating Poisson process rate, $\Omega$. All four plots show results of our MCMC sampler in black. Timing results for the matrix exponentiation method are represented by a purple horizontal line because the matrix exponentiation result does not vary as a function of $\Omega$. The two plots in the top row show results for a randomly generated tree with 50 tips. The two plots in the bottom row show results for a randomly generated tree with 100 tips. The rate matrix for the plots in the left column had 4 states. The rate matrix for the plots in the right column had 60 states.}
\label{timeVomega}
\end{figure}

Our MCMC approach scales well beyond state spaces of size 60 though matrix exponentiation does not. Timing results for our MCMC approach at larger state space sizes can be found in Figure D-1 of the Supplementary Materials. Matrix exponentiation-based stochastic mapping was prohibitively slow on state spaces reported in Figure D-1.

\subsection{Effect of the Dominating Poisson Process Rate}

Our tuning parameter, the dominating Poisson process rate $\Omega$, balances speed against mixing for our MCMC approach. The larger $\Omega$ is the slower the MCMC 
runs and the better it mixes. The optimal value for $\Omega$ depends on the CTMC state space and on the entries of the CTMC rate matrix. 
In our experience, it is not difficult to find a reasonable value for $\Omega$ for a fixed tree and a fixed rate matrix by trying different $\Omega$ values. We show the results of this exploration in Figure \ref{timeVomega} for two different values of the state space size, 4 and 60, and for two different trees, with 50 and 100 tips.

The top left plot in Figure \ref{timeVomega} shows the balance between speed and mixing most clearly. The optimal value for $\Omega$ appears to be around $0.2$ for 4 states and 50 tips. Our MCMC approach is clearly faster than the matrix exponentiation approach for a wide range of $\Omega$ values when the size of the state space is 60. When the size of the state space is 4 the matrix exponentiation approach can be faster, which is not surprising given the small size of the state space. Matrix exponentiation is about two times faster than our MCMC approach for the 100 tip tree with 4 states. Our MCMC approach can yield comparable speeds to the matrix exponentiation approach for the 50 tip tree with 4 states.

\subsection{Effect of Sparsity}
\begin{figure}[!htbp]
\begin{center}
\includegraphics[width=.9\textwidth]{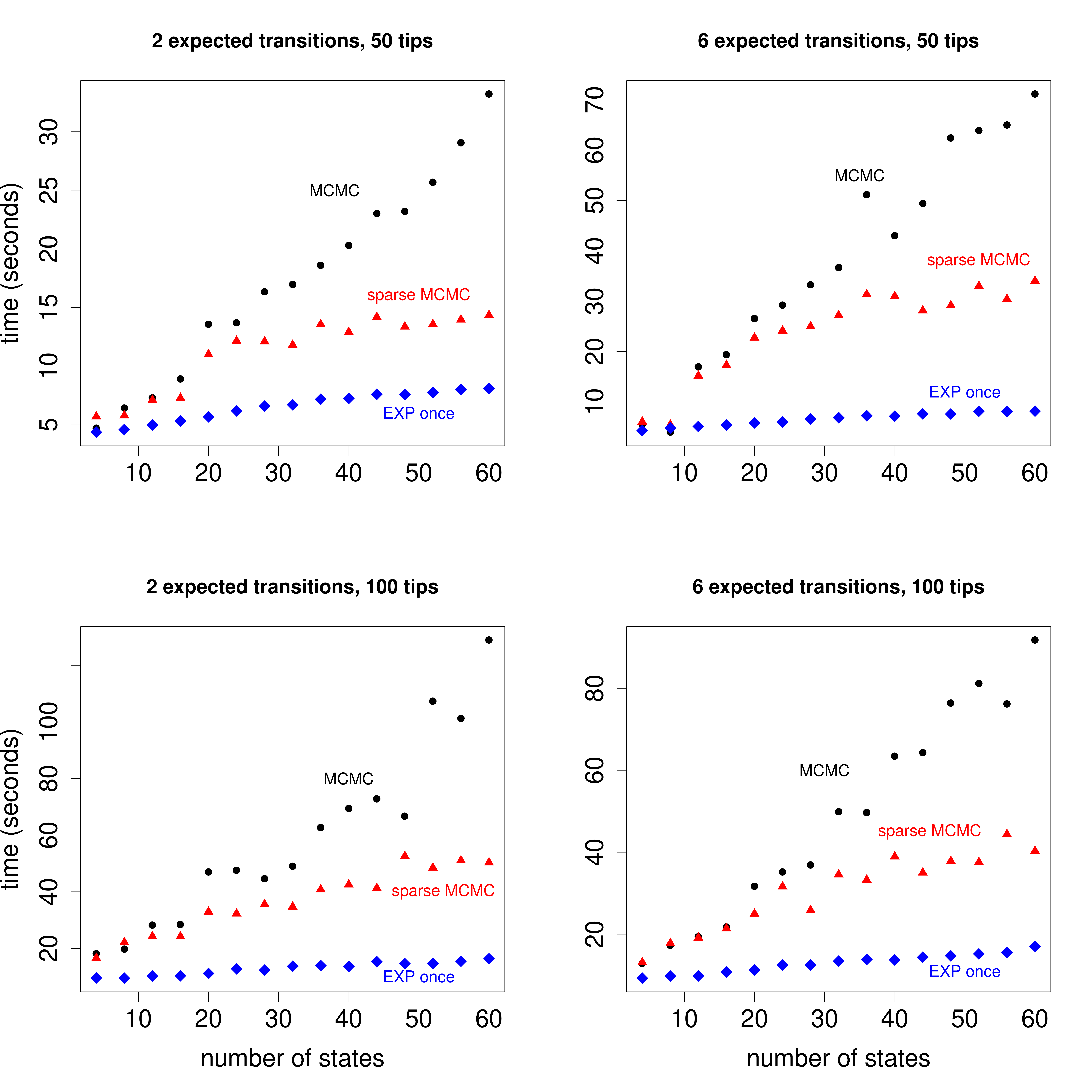}
\end{center}
\vspace{-0.4cm}
\caption{Time to obtain 10,000 effective samples as a function of the size of the state space. All four plots show results for three different implementations, our MCMC sampler in black, a sparse version of our MCMC sampler in red, and a matrix exponentiation approach that only exponentiates the rate matrix once per branch in blue. The rate matrix is tridiagonal and scaled to produce 2 expected transitions per tree (in the left column) or 6 expected transitions per tree (in the right column). The two plots in the top row show results for a randomly generated tree with 50 tips. The two plots in the bottom row show results for a randomly generated tree with 100 tips. The dominating Poisson process rate, $\Omega$, is 0.2.}
\label{sparse}
\end{figure}

Unlike matrix exponentiation methods, our new MCMC sampler is able to take advantage of sparsity in the CTMC rate matrix. There are three steps in our algorithm that can take advantage of sparsity: computing the partial likelihood matrix, sampling internal node states, and resampling branch states. In all three situations we need to multiply
 ${\bf B}^{M}$ by a vector of length $s$ -- the size of the state space. For a dense matrix this takes $\mathcal{O}(Ms^2)$ operations. When matrix $\mathbf{B}$ is sparse, the
 above multiplication requires fewer operations. For example, multiplying a vector by 
 $\mathbf{B}^M$ takes $\mathcal{O}(Ms)$ operations when $\mathbf{B}$ is triadiagonal. 
 It is interesting to note that while matrix exponentiation approaches cannot take advantage of sparsity when creating the partial likelihood matrix they can use sparsity when sampling branches via the uniformization technique of \citet{lartillot2006conjugate}.  

Speed increases due to sparsity depend on the size of the state space and the degree of sparsity in the probability transition matrix, ${\bf B}$.  In Figure \ref{sparse} we 
contrast the sparse implementation of our MCMC method with the
 implementation that does not take advantage of sparsity. 
 Figure \ref{sparse} also shows timing results for a matrix exponentiation method that only exponentiates the rate matrix once. 
\par
For a state space of size 60, the sparse implementation is about 2 times faster than the non-sparse implementation. Exponentiating the rate matrix once was always faster than the sparse implementation, sometimes by a factor of 4. We used uniformization to sample substitution histories for individual branches within the matrix exponentiation algorithm. This portion of the algorithm can take advantage of sparsity but there was not a large overall difference in run times between the sparse and non-sparse implementations. 

\subsection{Models of Protein Evolution}
\begin{figure}
\begin{center}
\includegraphics[width=.9\textwidth]{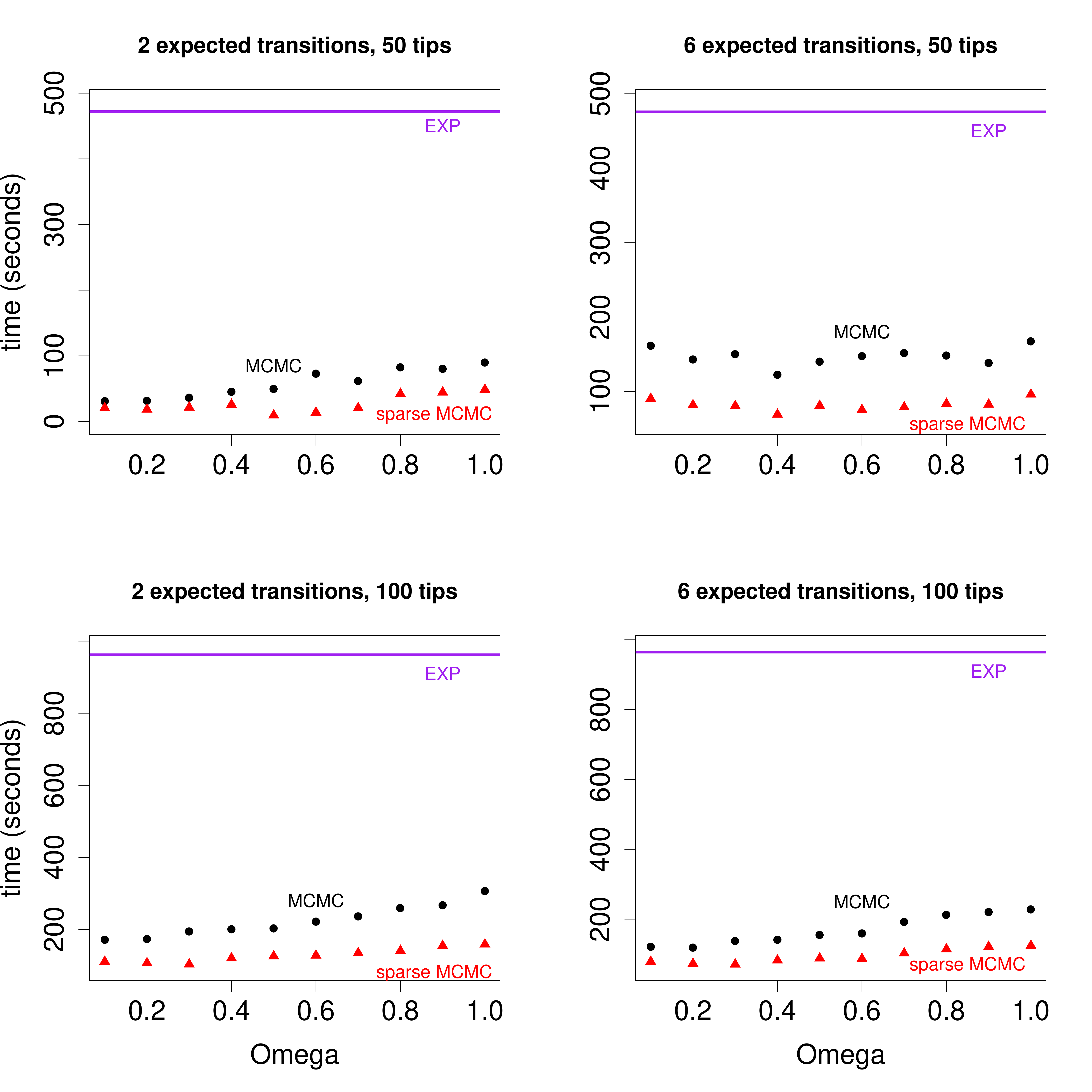}
\end{center}
\vspace{-0.5cm}
\caption{Time to obtain 10,000 effective samples as a function of the dominating Poisson process rate, $\Omega$, for the GY94 codon rate matrix. All four plots show results for three different implementations: our MCMC sampler in black, a sparse version of our MCMC sampler in red, and a matrix exponentiation approach in purple. 
The GY94 rate matrix was scaled to produce 2 expected transitions per tree (in the left column) or 6 expected transitions per tree (in the right column). The two plots in the top row show results for a randomly generated tree with 50 tips. The two plots in the bottom row show results for a randomly generated tree with 100 tips.}
\label{codon}
\end{figure}

We now turn to the investigation of efficiency of our new phylogenetic stochastic mapping in the context of modeling protein evolution. 
Evolution of protein coding sequences can be modeled on the following state spaces: state space of 4 DNA bases/nucleotides, 
state space of 20 amino acids, and state space of 61 codons --- nucleotide triplets --- excluding the three stop codons. 
The codon state space is the most computationally demanding of the three, causing existing phylogenetic mapping approaches to slow down considerably. 
The increased complexity that comes from modeling protein evolution at the codon level enables investigations into selective pressures and makes efficient use of the phylogenetic information for phylogeny reconstruction \citep{ren2005empirical}. 

In our numerical experiments, we use the Goldman-Yang-94 (GY94) model --- a popular codon substitution model proposed by \citet{goldman1994codon}, where the rate of substitution between codons depends on whether the substitution is synonymous (the codon codes for the same amino acid before and after the substitution) or nonsynonymous and whether the change is a transition 
($A\leftrightarrow G$, $C\leftrightarrow T$) or a transversion. The rate matrix is parameterized by  a synonymous/nonsynonymous rate ratio,
$\omega$, a transition/transversion ratio, $\kappa$,  and a stationary distribution of the CTMC, $\boldsymbol{\pi}^c$. 
The non-diagonal entries of the GY94 rate matrix, as described  are
\begin{align*}
q_{ab}&=\left\{\begin{array}{ll}
\omega \kappa \pi^c_b&\text{ if } a \rightarrow b \text{ is a non-synonymous transition},\\
\omega \pi^c_b&\text{ if } a \rightarrow b \text{ is a non-synonymous transversion},\\
\kappa \pi^c_b&\text{ if } a \rightarrow b \text{ is a synonymous transition},\\
\pi^c_b&\text{ if } a \rightarrow b \text{ is a synonymous transversion},\\
0&\text{ if $a$ and $b$ differ by 2 or 3 nucleotides}.
\end{array}\right. 
\end{align*}
The diagonal rates are determined by the fact that the rows of $\mathbf{Q}$ must sum to zero. 
In our simulations, we used the default GY94 rate matrix as found in the phylosim R package \citep{sipos2011phylosim}. 
The dominating Poisson process rate, our tuning parameter $\Omega$, ranged between being 8 times larger than the largest rate of leaving a state to being 80 times larger.  Timing results for the GY94 codon model rate matrix can be found in Figure \ref{codon}. GY94 contains structural zeros allowing our MCMC sampler to take advantage of sparsity and improve running times. Our MCMC approach was about 5 times faster than exponentiating the rate matrix at each iteration. A sparse version of our MCMC approach was about ten times faster than the matrix exponentiation 
method. 
\par
Encouraged by the computational advantage of our method on the codon state space, we also compared our new algorithm and 
the matrix exponentiation method on the amino acid state space. We used an amino acid substitution model called JTT, proposed by \citet{jones1992rapid}. The results can be found in Figure B-1 of the Supplementary Materials. We found that our MCMC approach 
is competitive even on the amino acid state space, but does not clearly outperform the matrix exponentiation method. This finding is not 
surprising in light of the fact that the size of the amino acid state space is three times smaller than the size of the codon state space.



\section{Discussion}

We have extended the work of \citet{rao2012fast} on continuous time HMMs to phylogenetic stochastic mapping. Our new method avoids matrix exponentiation, an operation that all current state-of-the-art methods rely on. There are two advantages to avoiding matrix exponentiation: 1) matrix exponentiation is computationally expensive for large CTMC 
state spaces; 2) matrix exponentiation can be numerically unstable. In this manuscript, we concentrated on the former advantage, because it is easier to quantify. 
However, it should be noted that numerical stability of matrix exponentiation is an obstacle faced by all phylogenetic inference methods. 
Currently, the most popular approach is to employ a reversible CTMC model, whose infinitesimal generator is similar to a symmetric matrix and therefore, can be robustly exponentiated via eigendecomposition \citep{Schabauer2012}.   
Researchers typically shy away from non-reversible CTMC models, to a large extent, because of instability of the matrix exponentiation of these models' 
infinitesimal generators \citep{Lemey2009bayesian}. In our new approach to phylogenetic stochastic mapping, we do not rely on properties of reversible CTMCs, making our 
method equally attractive for reversible and nonreversible models of evolution.
\par
We believe our new method will be most useful when integrated into a larger MCMC targeting a joint distribution of phylogenetic tree topology, branch lengths, and substitution
model parameters. 
Our optimism stems from the fact that stochastic mapping has already been successfully used in this manner in the context of complex models of protein evolution
\citep{lartillot2006conjugate,  rodrigue2008uniformization}. These authors alternate between using stochastic mapping to impute unobserved substitution histories and updating model
parameters conditional on these histories.  
We plan to incorporate our new MCMC algorithm into a conjugate Gibbs framework of \citet{lartillot2006conjugate} and \citet{rodrigue2008uniformization}. 
Since such a MCMC algorithm will operate on the state space of augmented substitution histories and model parameters, replacing Monte Carlo with MCMC in phylogenetic stochastic mapping may have very little impact on the overall MCMC mixing
and convergence. A careful study of properties of this new MCMC will be needed to justify this claim.


The computational advances made in \citep{lartillot2006conjugate,  rodrigue2008uniformization} are examples of 
considerable research activity aimed at speeding up statistical inference under complex models of protein evolution, 
prompted by the emergence of large amounts of sequence data \citep{Lartillot2013, Valle2014}. 
Challenges encountered in these applications also appear in statistical applications of many other models of evolution
that operate on large state spaces: models of microsatellite evolution \citep{Wu2011},
models of gene family size evolution \citep{Spencer2006}, phylogeography models \citep{Lemey2009bayesian}, 
and covarion models \citep{Penny2001, galtier2001maximum}.
Our new phylogenetic stochastic mapping without matrix exponentiation should be a boon for researchers using these models 
and  should enable new analyses that, until now, were too computationally intensive to be attempted.

\section*{Acknowledgments}
We thank Jeff Thorne and Alex Griffing for helpful discussions and for their feedback on an early version of this manuscript and Jane Lange
for pointing us to the work of \citet{rao2012fast}. 
VNM was supported in part by the National Science Foundation grant DMS-0856099 and by the National Institute of Health grant R01-AI107034.
JI was supported in part by the UW NIGMS sponsored Statistical Genetics Training grant\# NIGMS T32GM081062.

\bibliography{no_mat_exp}

\clearpage

\renewcommand{\thefigure}{A-\arabic{figure}}
\setcounter{figure}{0}

\section*{Appendix A}

We present evidence supporting the claim that the stationary distribution of our new MCMC sampler is the posterior distribution, $p(\mathcal{V},\mathcal{W}|{\bf Y})$. This posterior has many aspects that could be examined, but for simplicity we focus on univariate statistics: the amount of time spent in each state and the number of transitions between each pair of states.

 We compare the results of five different implementations. The first is a sampler implemented in the diversitree package \citep{fitzjohn2012diversitree}, labeled diversitree or DIV. The second is our version of the same method, labeled EXP. The third is the same method that only exponentiates the rate matrix once, labeled EXP ONCE or ONCE. The fourth is our new method, labeled MCMC. The fifth is a sparse version of our new method, labeled SPARSE or SPA. 

We present results for four regimes, two different sizes of state spaces, and two different sets of transition rates. The smaller state space has 4 states and the larger has 20. The lower transition rates correspond to 2 expected transitions per tree and the higher transition rates correspond to 20 expected transitions per tree. In an effort to reduce the number of plots we focus only on states that were observed at the tips of the tree. All four simulated trees had 20 tips.

\begin{figure}[!h]
\begin{center}
\includegraphics[width=120mm]{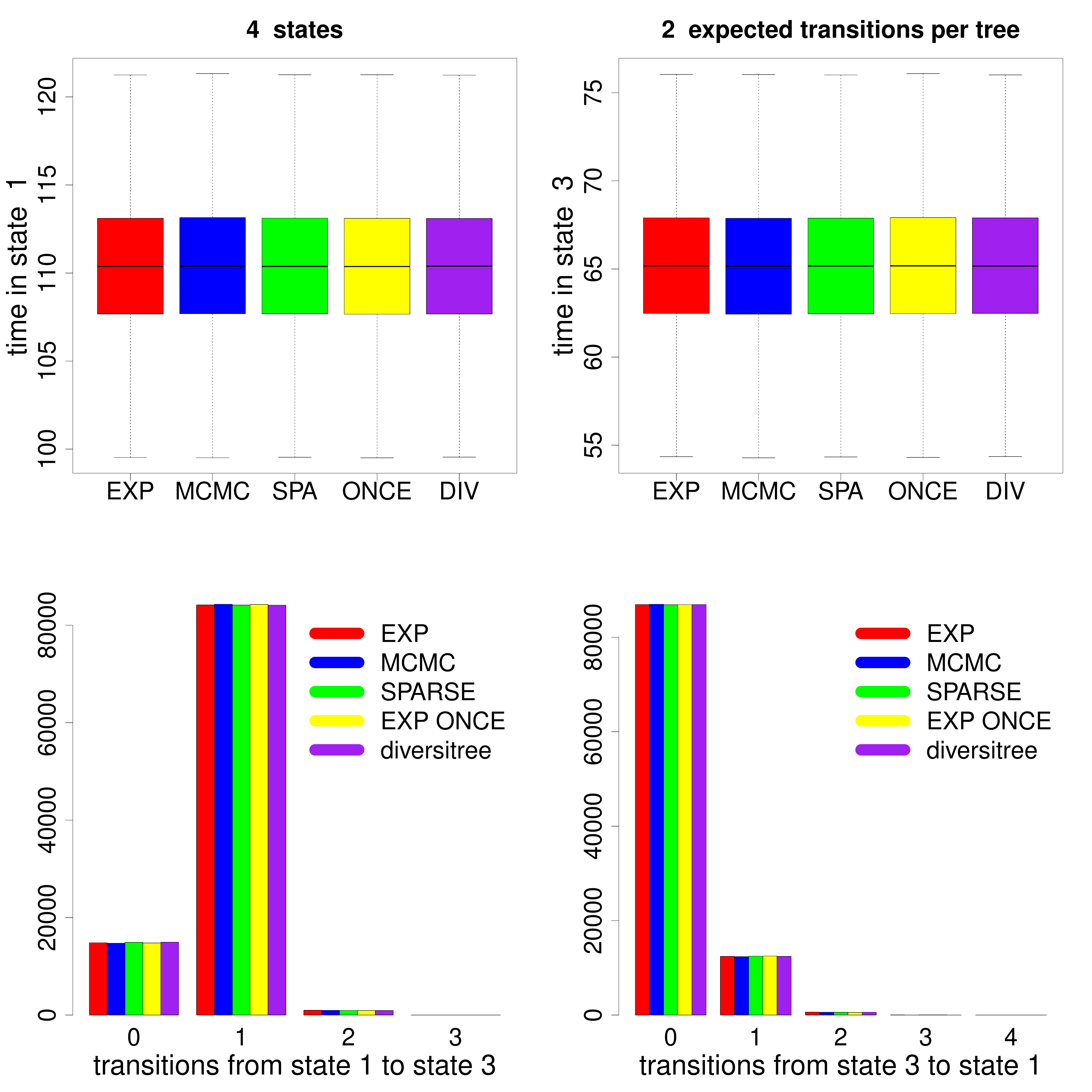}
\end{center}
\caption{Univariate summaries for five implementations of state history sampling of a 20 tip tree. There were 4 states and 2 expected transitions per tree. The top plots contain boxplots illustrating the distribution of the amount of time spent in state 1 and state 3. Outliers were not included though all five implementations showed the same outlier behavior. The bottom plots contain histograms illustrating the posterior distribution of the number of transitions between state 1 and state 3.}
\label{app1}
\end{figure}

Our first example used the smaller state size, 4, with the smaller number of expected transitions per tree, 2. A random simulation resulted in two unique tip states, states 1 and 3. For each method we produced 100,000 state history samples. Figure \ref{app1} contains plots of four univariate statistics pulled from the posterior distributions. All five implementations produced the same results.

\begin{figure}
\begin{center}
\includegraphics[width=120mm]{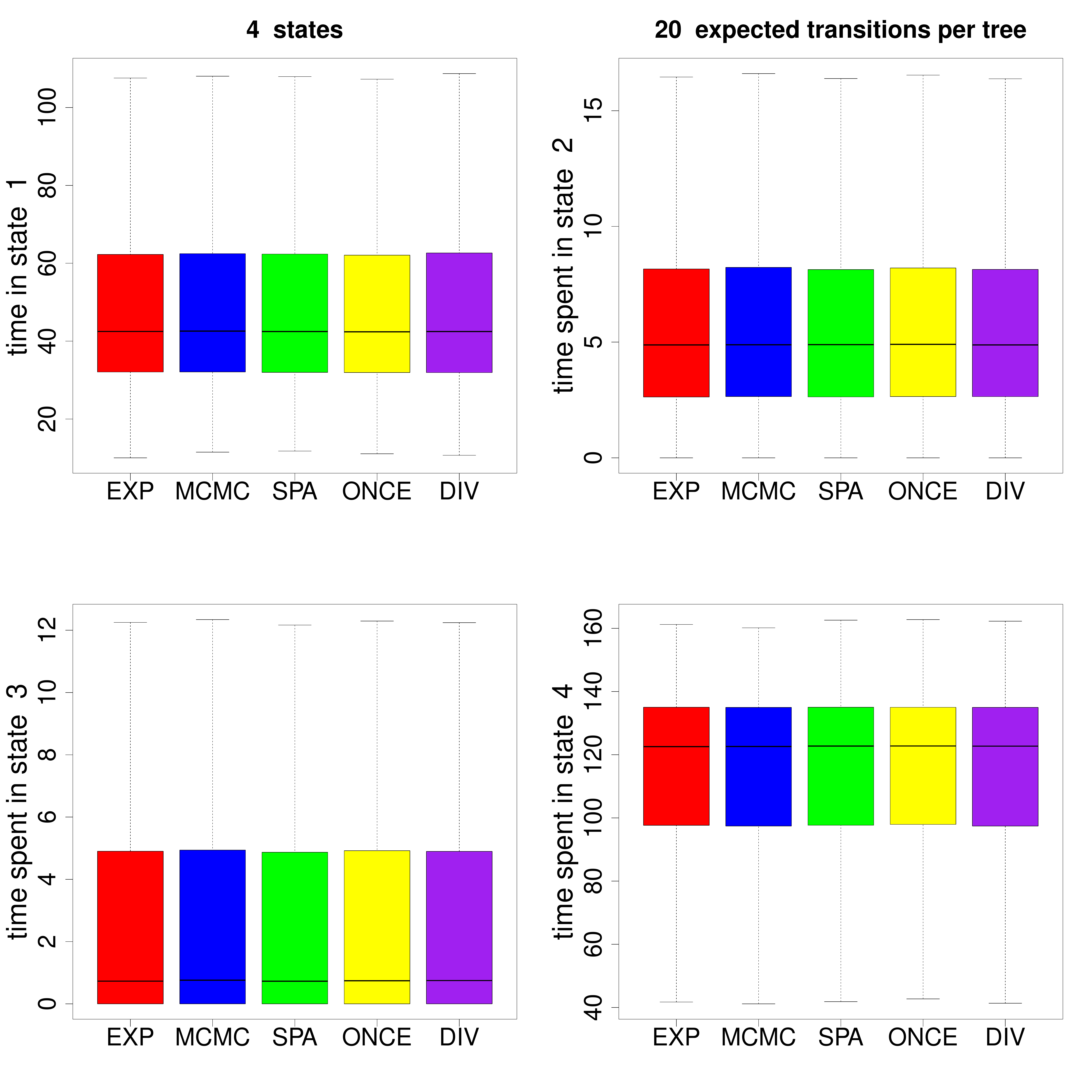}
\end{center}
\caption{Boxplots illustrating the posterior distribution of the amount of time spent in each state. Outliers were not included though all five implementations showed the same outlier behavior. There were 4 states and 20 expected transitions per tree.}
\label{app2}
\end{figure}

\begin{figure}
\begin{center}
\includegraphics[width=120mm]{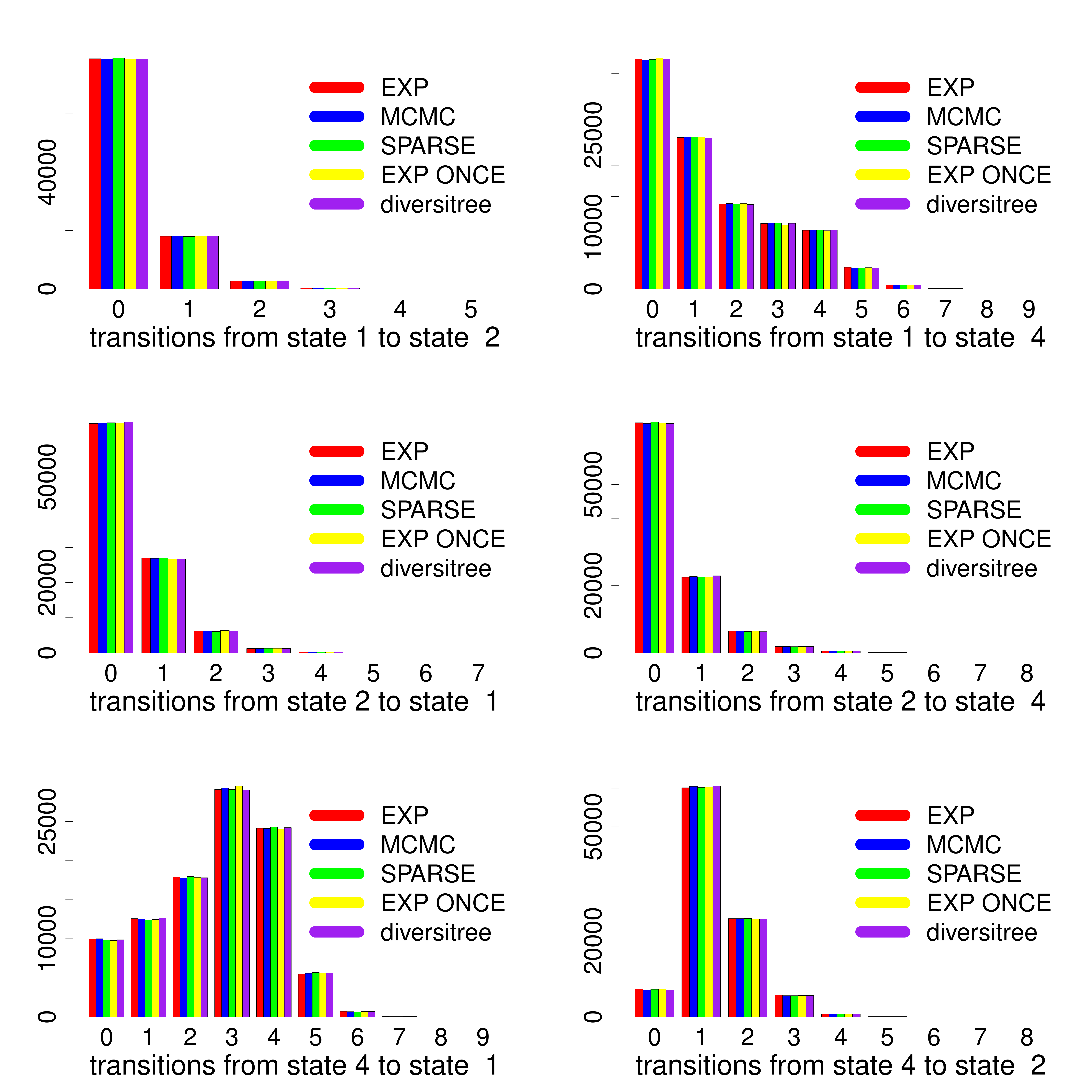}
\end{center}
\caption{Histograms illustrating the posterior distribution of the number of transitions between states 1, 2, and 4. There were 4 states and 20 expected transitions per tree. }
\label{app3}
\end{figure}

Our second example used the smaller state size, 4, with the larger number of expected transitions per tree, 20. A random simulation resulted in three unique tip states, states 1, 2, and 4. Figure \ref{app2} contains four boxplots pulled from the posterior distributions. Figure \ref{app3} contains six histograms pulled from the posterior distributions. 

\begin{figure}
\begin{center}
\includegraphics[width=120mm]{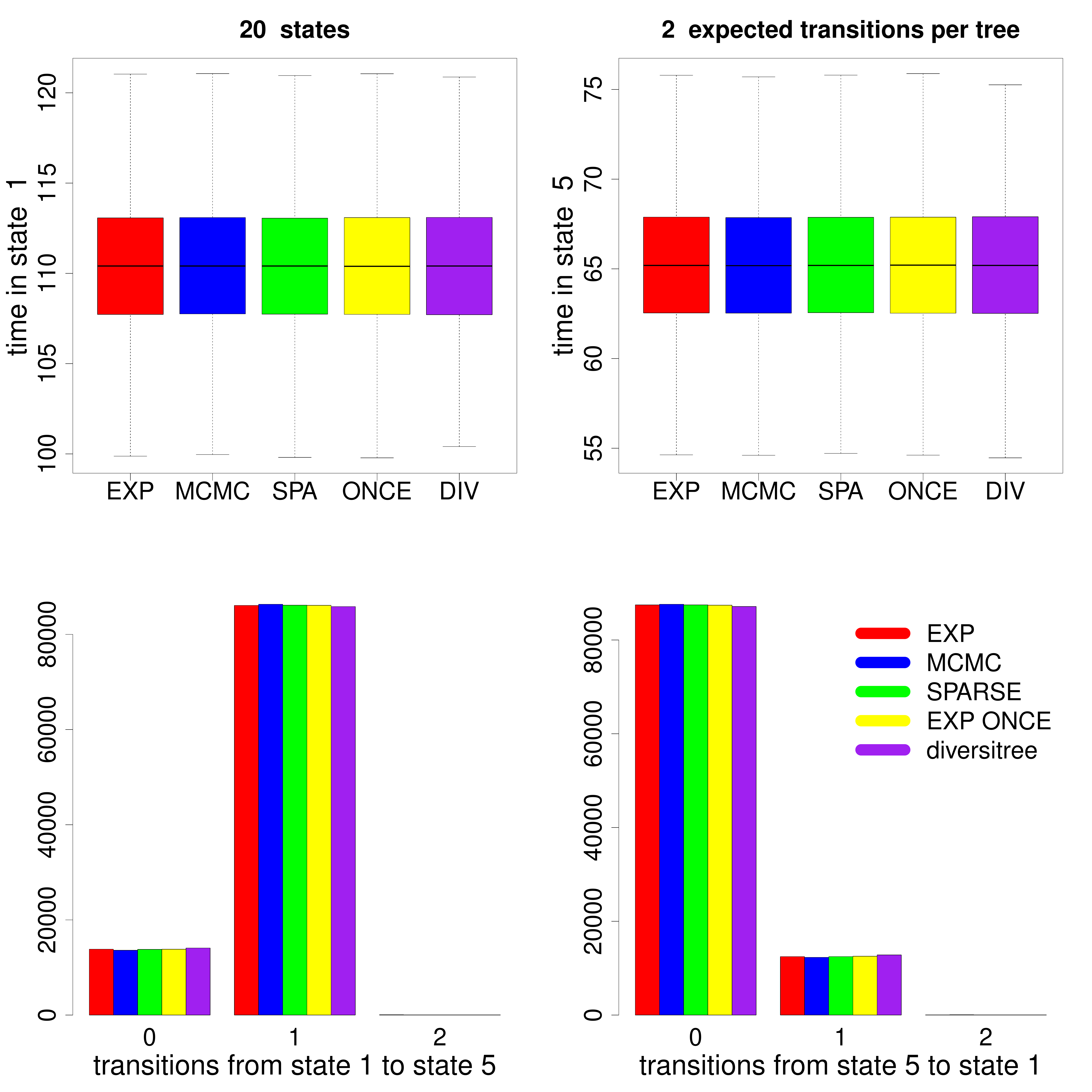}
\end{center}
\caption{Univariate summaries for 5 implementations of state history sampling of a 20 tip tree. There were 20 states and 2 expected transitions per tree. The top plots contain boxplots illustrating the posterior distribution of the amount of time spent in state 1 and state 5. Outliers were not included though all five implementations showed the same outlier behavior. The bottom plots contain histograms illustrating the posterior distribution of the number of transitions between state 1 and state 5.}
\label{app4}
\end{figure}

Our third example used the larger state size, 20, with the smaller number of expected transitions per tree, 2. A random simulation resulted in two unique tip states, states 1 and 5.  Figure \ref{app4} contains plots of four univariate statistics pulled from the posterior distributions

\begin{figure}
\begin{center}
\includegraphics[width=120mm]{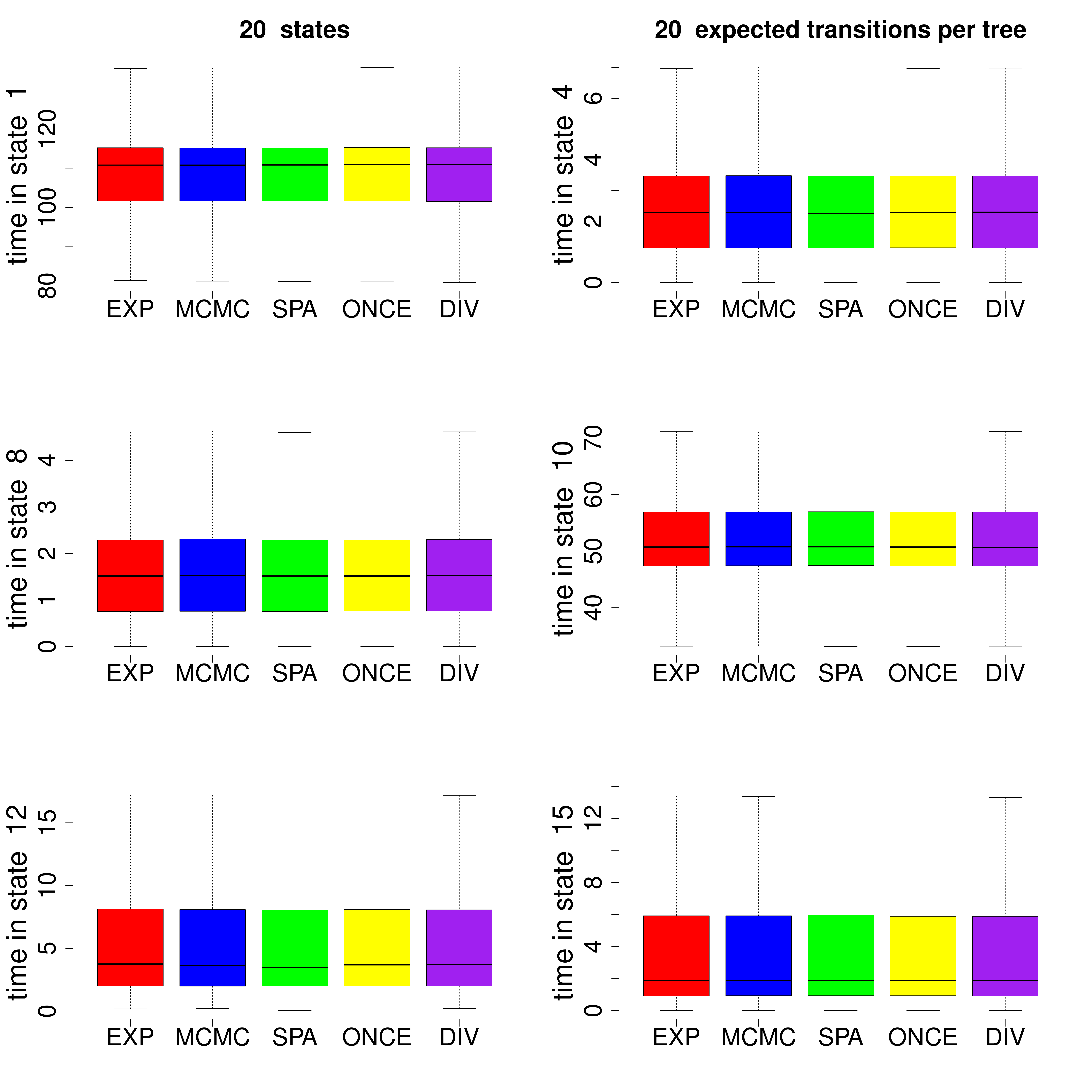}
\end{center}
\caption{Boxplots illustrating the posterior distribution of the amount of time spent in each tip state. Outliers were not included though all five implementations showed the same outlier behavior. There were 20 states and 20 expected transitions per tree. }
\label{app5}
\end{figure}

\begin{figure}
\begin{center}
\includegraphics[width=120mm]{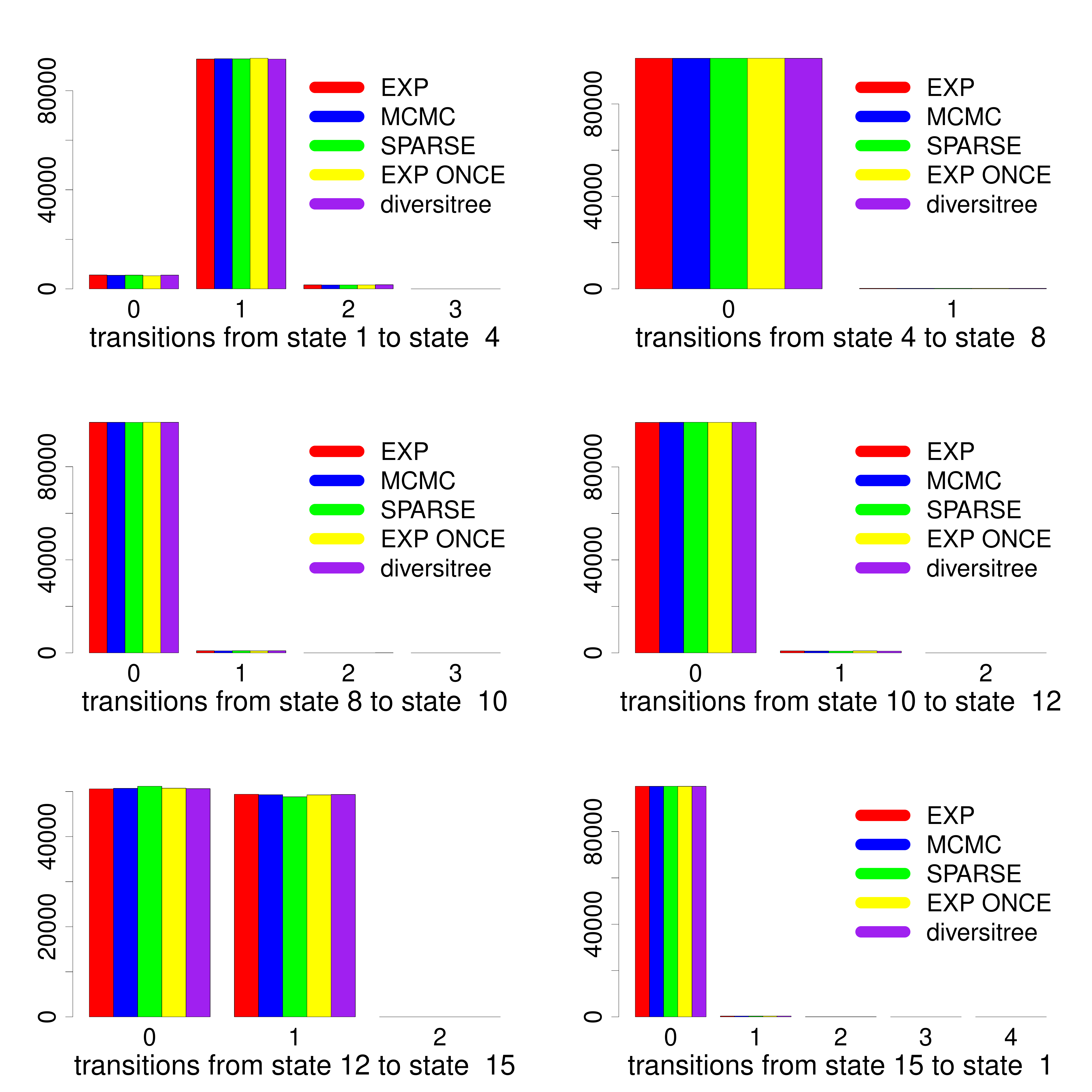}
\end{center}
\caption{Histograms illustrating the posterior distribution of the number of transitions between a subset of the tip states. There were 20 states and 20 expected transitions per tree. }
\label{app6}
\end{figure}

Our fourth example used the larger state size, 20, with the larger number of expected transitions per tree, 20. A random simulation resulted in six unique tip states, states 1, 4, 8, 10, 12, and 15. Figure \ref{app5} contains six boxplots pulled from the posterior distributions. Figure \ref{app6} contains six histograms pulled from the posterior distributions. 

\renewcommand{\thefigure}{B-\arabic{figure}}
\setcounter{figure}{0}
\section*{Appendix B}

\begin{figure}
\begin{center}
\includegraphics[width=.9\textwidth]{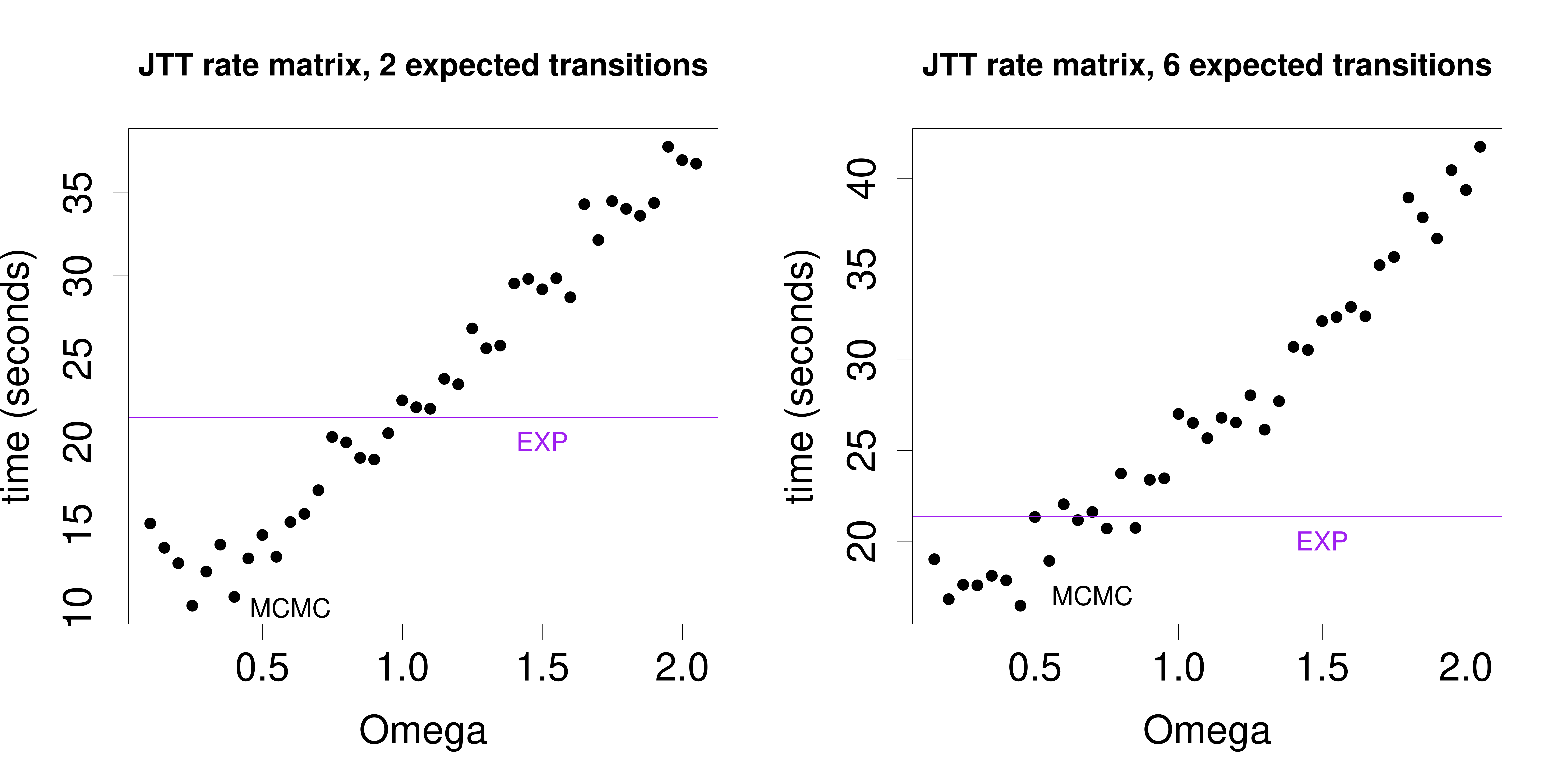}
\end{center}
\caption{Time to obtain 10,000 effective samples as a function of the dominating Poisson process rate, $\Omega$, for the JTT amino acid rate matrix as found in the phylosim R package. Results for our MCMC sampler are shown in black. Timing results for the matrix exponentiation method are represented by a purple horizontal line because the matrix exponentiation result does not vary as a function of $\Omega$. The randomly generated tree had 40 tips. The JTT rate matrix was scaled to produce 2 expected transtions in the left hand plot. The JTT rate matrix was scaled to produce 6 expected transtions in the right hand plot.}
\label{amino}
\end{figure}

One state space of interest in molecular evolution is the amino acid state space. \citet{jones1992rapid} proposed a rate matrix for an amino acid CTMC substitution model, called JTT. Figure \ref{amino} contains timing results for this rate matrix and a tree with 40 tips. When our MCMC approach used an appropriately tuned value of $\Omega$ we saw slightly faster running times as compared to the matrix exponentiation approach. We examined other scenarios for the JTT rate matrix in which we saw faster running times with the matrix exponentiation approach.

\renewcommand{\thefigure}{C-\arabic{figure}}
\setcounter{figure}{0}
\section*{Appendix C}

\begin{figure}
\begin{center}
\includegraphics[width=0.8\textwidth]{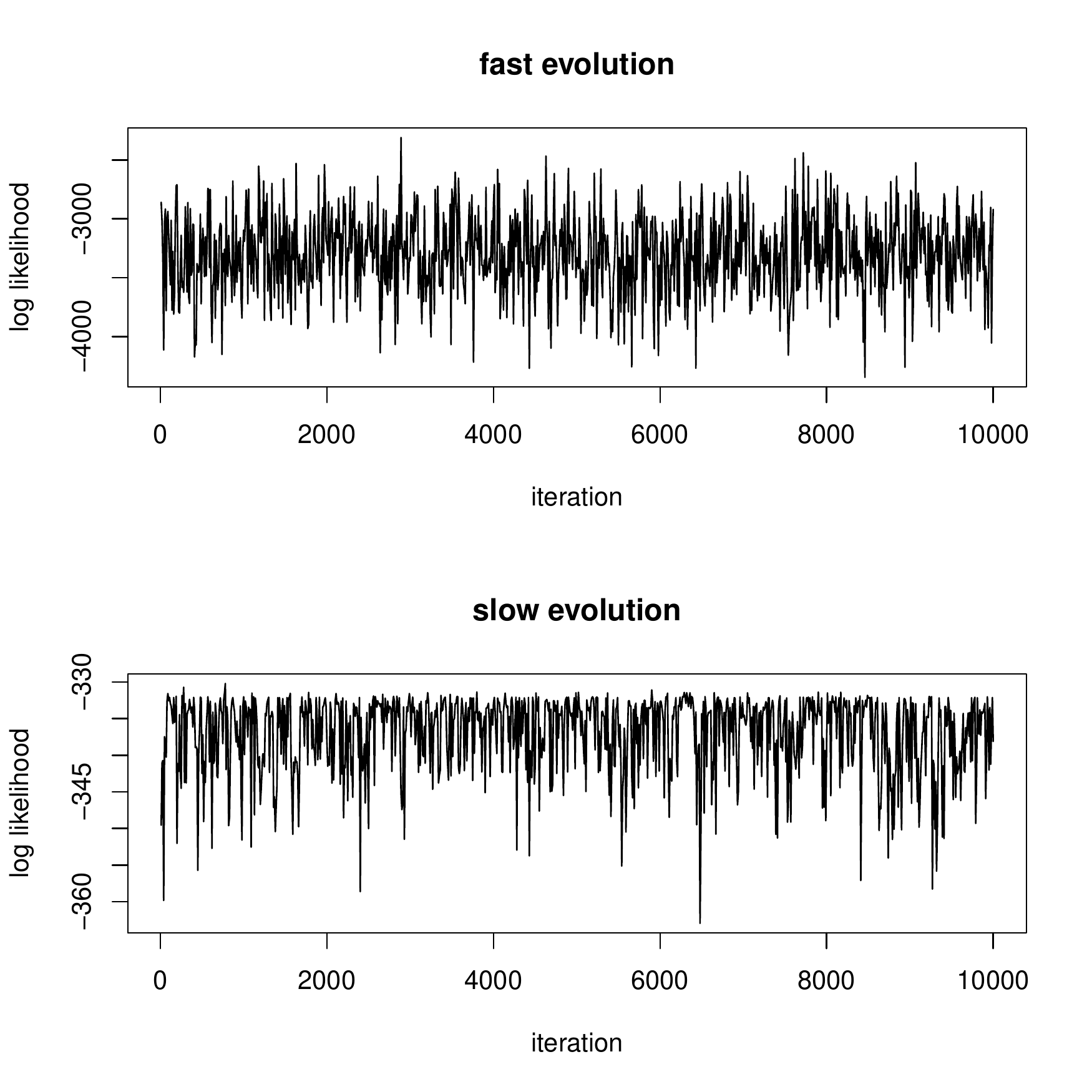}
\end{center}
\caption{MCMC trace plots. We show the log density of substitution histories for two MCMC chains at every tenth iteration. The top plot shows results for a trait that evolved quickly (with 
6 expected substitutions). The bottom plot shows results for a trait that evolved slowly (with 2 expected substitutions). 
In both cases, the tree had 50 tips and the size of the state space was 10.}
\label{convergence}
\end{figure}

Our MCMC sampler seems to converge to stationarity quickly. Figure \ref{convergence} shows two convergence plots, one for fast evolution and one for slow evolution. In both cases we started the chain with an augmented substitution history containing one transition in the middle of each branch leading to a tip whose state was different from an arbitrarily chosen root state. In the case of slow evolution this substitution history was a poor starting point but the log likelihood of the chain appeared to achieve stationarity quickly. In both cases, the tree had 50 tips and the size of the state space was 10. 

\renewcommand{\thefigure}{D-\arabic{figure}}
\setcounter{figure}{0}
\section*{Appendix D}

\begin{figure}
\begin{center}
\includegraphics[width=0.8\textwidth]{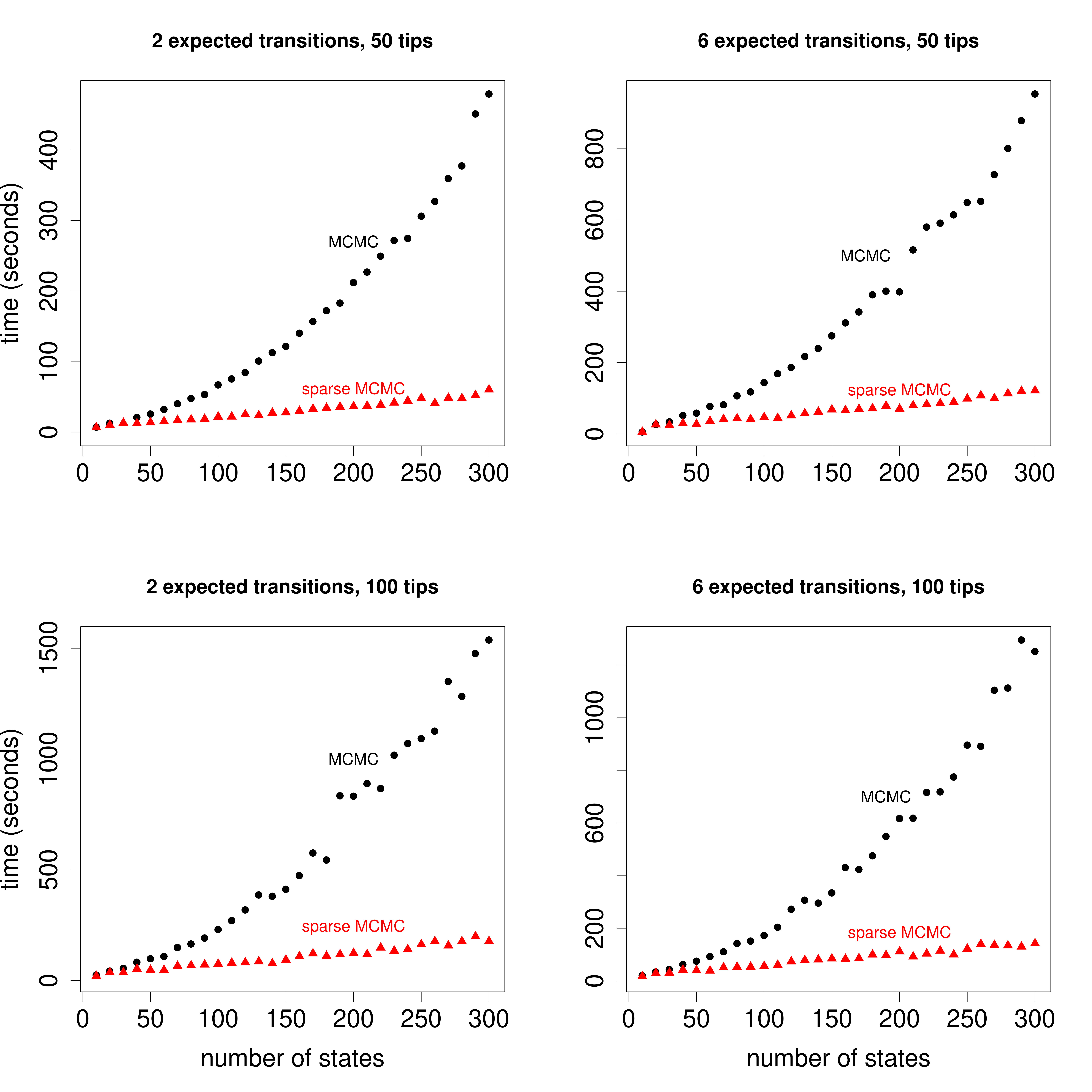}
\end{center}
\caption{State space effect for a tridiagonal rate matrix. All four plots show the amount of time required to obtain 10,000 effective samples as a function of the size of the state space for two methods, our MCMC sampler in black circles and a sparse version of our MCMC sampler in red triangles. The two plots in the top row show results for a randomly generated tree with 50 tips. The two plots in the bottom row show results for a randomly generated tree with 100 tips. The two plots in the left column show results for a rate matrix that was scaled to produce 2 expected transitions while the two plots in the right column show results for a rate matrix that was scaled to produce 6 expected transitions.}
\label{largerstates}
\end{figure}

Our MCMC method scales well with the size of the state space even when state space sizes exceed 100. Figure \ref{largerstates} shows timing results for state space sizes going out to 300. We show results for our MCMC method and a sparse version of our MCMC method using tridiagonal rate matrices. 

\end{document}